\documentclass[conference]{IEEEtran}
\ifCLASSINFOpdf
\else
\fi
\usepackage{amsfonts}
\usepackage{bbm}
\usepackage{mathrsfs}
\usepackage{cite}
\usepackage{times}
\usepackage{graphicx}
\usepackage{amsfonts}
\usepackage{flushend}
\usepackage{algorithmic}
\usepackage[ruled]{algorithm}
\usepackage{cite}
\usepackage{times}
\usepackage{amsmath}
\usepackage{graphicx}
\usepackage{epsfig}
\usepackage{amsmath}
\usepackage{latexsym}
\usepackage{amsfonts}
\usepackage{amssymb}
\usepackage{paralist}
\usepackage{xspace}
\usepackage{mathrsfs}
\usepackage{amssymb}
\usepackage{color}
\usepackage{algorithmic}
\usepackage[ruled]{algorithm}
\usepackage{flushend}
\usepackage{subfigure}
\usepackage{makecell}
\usepackage{multirow}

\usepackage{slashbox,pict2e}
\usepackage{calc}
\usepackage{tikz}

\newtheorem{defn}{Definition~}

\def\ie{\textit{i.e.}\xspace}
\def\etal{\textit{et al.}\xspace}
\def\etc{\textit{etc.}\xspace}
\def\eg{\textit{e.g.}\xspace}

\renewcommand{\baselinestretch}{0.98}
\begin{document}
%

\title{Improving Data Forwarding in Mobile Social Networks with Infrastructure Support: \\
A Space-Crossing Community Approach}

\author{\authorblockN{Zhong Li\authorrefmark{1}, Cheng Wang\authorrefmark{2}, Siqian Yang\authorrefmark{1}, Changjun Jiang\authorrefmark{1}, Ivan Stojmenovic\authorrefmark{2}
}
\authorblockA{\authorrefmark{1} Department of Computer
Science, Tongji University, Shanghai, China}
\authorblockA{\authorrefmark{2} School of Information Technology and
Engineering, University of Ottawa, Canada}

}

\maketitle

\begin{abstract}

In this paper, we study two tightly coupled issues: space-crossing community detection and its influence on data
forwarding in Mobile Social Networks (MSNs) by taking the
hybrid underlying networks with infrastructure support into consideration.
The hybrid underlying network is composed of large numbers of mobile users and a small portion of Access Points (APs).
Because APs can facilitate the communication
among long-distance nodes, the concept of physical proximity community  can be extended to be one across the geographical space.
In this work, we first
investigate  a space-crossing community detection method for MSNs.
Based on the detection results,
 we design  a novel data forwarding algorithm SAAS (Social Attraction and AP Spreading), and show how to
exploit the space-crossing communities to improve the data forwarding efficiency.
We evaluate our SAAS algorithm on real-life data from
MIT Reality Mining and UIM.
Results show that space-crossing community plays a positive role in data forwarding in MSNs in terms of deliver ratio and delay.  Based on this new type of community, SAAS achieves a better performance than existing
social community-based data forwarding algorithms in practice, including Bubble Rap and Nguyen's Routing algorithms.
\end{abstract}

\IEEEpeerreviewmaketitle

\section{Introduction}

As the development of social networks, more and more Bluetooth/WiFi-based mobile social applications emerge, for example, MobiClique\cite{pietilainen2009mobiclique}, Foursquare \cite{Foursquare}, E-SmallTalker\cite{yang2010smalltalker}, Sony's Vita\cite{sony}. These applications need real-time communication ensuring. Traditional method of deploying large numbers of base stations costs too much and the work load of the base stations is heavy. Subsequently, as an extended form of  centralized way, self-organized ad hoc underlying networks come into being. In ad hoc networks, there are no stable end-to-end delivery paths to guarantee a high efficient data forwarding. Therefore, many works have been studying for settling this problem. Recently,  solutions using social community detection results are proposed to design a data forwarding schemes in Mobile Social Networks (MSNs), such as \cite{hui2008bubble,gao2009multicasting,fan2012geo,overlap}.


 However, ad hoc networks mainly act as an extended form for centralized way rather than an individual network architecture in typical real-life applications. In reality, the underlying networks are usually hybrid with infrastructure support.
For example, in some areas, due to lack
of internet Access Points (APs) or very weak signals from cellular base
stations, users need to use a self-organized way, while in other areas, users can have APs to help them do some management. In this paper, we consider the hybrid underlying network with APs support and study how to utilize  APs to design a high efficient data forwarding scheme.

We consider a mobile social network in which several APs are deployed. The controlled areas of  APs \emph{cannot} cover the entire network and only occupy a small part. For simplicity, in an AP controlled area, users can access AP via one hop communication.
In such hybrid underlying network, APs facilitate the communication among some
long-distance nodes. Unlike the traditional concept of a community (A community is a group of tight-knit nodes with more internal than external links \cite{girvan2002community,porter2009communities,fortunato2010community}, such as the common interest community with logical links, the physical proximity community with geographic aggregation), we propose a novel concept of \emph{space-crossing community} which is a group of nodes with relatively stronger communication capability. Space-Crossing communities are gained through spanning physical space to merge  physical proximity communities and access point communities. The physical proximity community implies the strong capability of facilitating  communication among nodes in geography. The access point community forms a strong communication area itself. These communities and the combining among them will build our strong communication community, \ie, space-crossing community.

  Besides, we assume that each node $u$ has its \emph{local activity} associated with  a certain space-crossing community $ComSC$. It is the ratio of $u$'s encounter probability with other nodes in $ComSC$  to the sum of encounter probability between any two nodes in
$ComSC$. Note that the common concept of \emph{activity} is
with the whole network, not with a certain community of
a node. Node local activity reflects a statistics of encounter
probability in a node¡¯s certain space-crossing community.
To the best of our knowledge, this is the first paper that studies the space-crossing community detection and its influence on data forwarding in MSNs by taking the hybrid underlying network with APs support into consideration. The main
contributions of this paper can be summarized as follows:

 $\bullet$ We give a space-crossing community detection method for MSNs, including the initializing phase and the dynamic tracking phase. Combination criterion $\mathcal{S}^{a}$, $\mathcal{S}^{b}$ and $\mathcal{S}^{c}$ work throughout the entire detection method to gain the final space-crossing communities.

 $\bullet$ We propose SAAS (Similarity Attraction and AP Spreading) data forwarding algorithm, in which we take full advantage of  the space-crossing communities. SAAS consists of two phases. In first phase, in non-AP areas, we use social similarity to guide data forwarding. Social similarity is defined by combining node local activity with pearson correlation coefficient. In second phase, in  AP controlled areas, we spread the message copies to make more nodes carry messages to the destination.

 $\bullet$ We extensively evaluate SAAS on two hybrid underlying networks: MIT Reality Mining \cite{uiuc-uim-2012-01-24} and UIM (University of Illinois Movement) \cite{eagle2009inferring} datasets. The results show that SAAS significantly outperforms several existing social community-based algorithms.

The rest of the paper is organized as follows. Section \ref{Network Model} presents the network model. Section \ref{space-crossing} and Section \ref{data forwarding} study the space-crossing community detection
and its impact on data forwarding in MSNs.  We conduct extensive
experiments and report our results in Section \ref{Evaluation}. We discuss several issues about our algorithm and experiments in Section \ref{Discussion}. We
review related work in Section \ref{Related Work} and conclude the paper in
Section \ref{Conclusion}.
\section{Network Model} \label{Network Model}
\subsection{Dynamic Graph}

Our mobile social network consists of  mobile users and stationary Access Points (APs). APs are deployed in the mobile social network originally, not being as supplementary devices for capacity improvement. We model this \emph{hybrid underlying network with APs support} as a \emph{dynamic graph} which can be defined as a time sequence of network graph, denoted by $\mathcal{G} =\{G_{0}, G_{1},...,G_{t},...\}$, where $G_{t}=(V_{t},E_{t})$ represents a time dependent network snapshot recorded at time $t$; $V_{t}$ denotes the set of nodes, including the set of mobile users and the set of stationary APs; $E_{t}=\{(u,v)|u,v\in V_{t}\}$ denotes the edge set. In terms of the entire network, both the node and edge sets change over time.

\subsection{Assumption of AP}
$\bullet$ An AP can be used as a relay or a centralized server.

$\bullet$ In our mobile social network, the ratio of the number of APs to the number of mobile users is small. Different from a base station, the coverage area and processing capability of an AP is limited, to be specific, the radius of coverage area is usually about 30-100m.

$\bullet$ The backbones among APs can be designed according to reality. As the first work on hybrid underlying networks with APs support, this paper makes a simple assumption that APs are connected in a circle, in order to highlight the main observation, i.e., the impact of infrastructure support on data forwarding in MSNs.
\subsection{Contact Aggregation for Edges}\label{Contact Aggregation}

The edge set $E_{t}$ is formed according to the following steps.

 1. When a mobile user enters into the controlled area of an AP, an edge will be formed between the mobile user and the AP.

2. For mobile users which are in a certain AP area, some edges are added among them to form a complete graph because of the strong connectedness of the AP.

3. For APs, some edges are added among them to form a circle.

4. From trace analysis of the real-life social datasets which  usually contain the contact records of bluetooth and wifi access points, we add the number of direct contacts between user pairs $u$ and $v$ iteratively in a chosen period $t_{0}$ to $t_{q}$. Denote the number of contacts between node $u$ and $v$ at time $t_{i}$ as $l_{uv}^{t_{i}}$; denote the number of contacts among all nodes in the network at time $t_{i}$ as $l_{\ast}^{t_{i}}$; denote the encouter ratio between node $u$ and $v$ at time $t_{i}$ as $w_{uv}^{t_{i}}$.

  $\bullet$ If the contact traces are sparse, we will implement a weighted growing window mechanism. Let $\sum\nolimits_{i=0}^{p} l_{uv}^{t_{i}}$ denote the overall numbers of contacts between user $u$ and $v$ in time period $t_{0}$ to $t_{p}$; let $\sum\nolimits_{i=0}^{p} l_{\ast}^{t_{i}}$  denote the overall  numbers of contacts for all users. Thus, we have an encounter ratio  value $w_{uv}^{t_{p}}=\frac{\sum\nolimits_{i=0}^{p} l_{uv}^{t_{i}}}{\sum\nolimits_{i=0}^{p} l_{\ast}^{t_{i}}}$ between user $u$ and $v$ at time $t_{p}$, where $0\leq p\leq q$.
    For any node $u^{'}$ and $v^{'}$, if $w_{uv}^{t_{p}}$ is larger than the median of $\{w_{u^{'}v^{'}}^{t_{p}} |u^{'},v^{'}\in V_{t} \}$, there will form an edge between user $u$ and $v$.

 $\bullet$ If the contact traces are dense, we will carry out a weighted sliding window mechanism. The time granularity of the window length $\Delta$ is empirically determined according to different datasets\footnote{Usually, the time granularity of the records in the dataset is very small (in second order). It is too short to  reflect the social properties and form social graphs. Thus, $\Delta$ is chosen at least larger than the time granularity of the dataset.}. Let $\sum\nolimits_{i=p-\Delta}^{p} l_{uv}^{t_{i}}$ denote the overall numbers of contacts between user $u$ and $v$ in last $\Delta$ time window length from the current time $t_{p}$; let $\sum\nolimits_{i=p-\Delta}^{p} l_{\ast}^{t_{i}}$  denote the overall  numbers of contacts for all users. Thus, we have an encounter ration   $w_{uv}^{t_{p}}=\frac{\sum\nolimits_{i=p-\Delta}^{p} l_{ij}^{t_{i}}}{\sum\nolimits_{i=p-\Delta}^{p} l_{\ast}^{t_{i}}}$ between user $u$ and $v$ at time $t_{p}$, where $\Delta\leq p\leq q$.
     For any node $u^{'}$ and $v^{'}$, if $w_{uv}^{t_{p}}$ is larger than the median of $\{w_{u^{'}v^{'}}^{t_{p}} |u^{'},v^{'}\in V_{t} \}$, there will will form an edge between user $u$ and $v$.

\emph{Remark 1}: In Step 4, in order to cope with the invalidation of bluetooth devices in those traces, we have $w_{uv}^{t_{p}}=w_{vu}^{t_{p}}$, with assigning the larger value for them.

\emph{Remark 2}: In Step 4, our \emph{weighted} aggregation method avoids the imperfection of the simple growing and sliding time window methods which will make a social graph more and more meshed to lose heterogeneity and degenerate to a random graph over time \cite{hossmann2010know}.

\emph{Remark 3}: It does not matter that some edges will be formed repeatedly in Step 2 and Step 4.
%

\subsection{Community Structure}

 The definition of community depends on the special community detection algorithms or social applications. I\emph{n our hybrid underlying network, generally, we say if people have relatively stronger communication capability with each other, they may form a community.} In our paper, three kinds of communities satisfy the condition of strong communication capability. The first two kinds (Physical Proximity Community and Access Point Community) are the components of the third kind (Space-Crossing Community). Here, we allow that communities can overlap with each other.

\emph{Physical Proximity Community}:
Due to physical proximity, if mobile users encounter with each other frequently in geography, according to the criterion that a community is a structure having a group of tight-knit nodes with more internal links than external links \cite{girvan2002community,porter2009communities,fortunato2010community}, they may form a Physical Proximity Community (PP Community). We will choose an appropriate community criterion among various methods, which is provided in Section \ref{FOCS}.

\emph{Access Point Community}: According to an AP controlled area, the mobile users and the AP in this area form an Access Point Community (AP Community).

\emph{Space-Crossing Community}: In our hybrid underlying network, some nodes are far away from each other in geography, but due to the help of APs, some long-distance nodes will have strong communication capability in the aspect of crossing the physical space. Thus, the long-distance nodes mixed with the close-proximity  nodes may  form  a Space-Crossing Community (SC Community).  We use the \emph{Space-Crossing Community Detection} algorithm  in Section \ref{space-crossing} to merge the Physical Proximity Communities and the Access Point Communities and gain the final Space-Crossing Communities.

\section{Space-Crossing Community Detection} \label{space-crossing}

In the dynamic environment, in order to handle the network changes quickly, we execute two steps to detect the space-crossing communities. At initial network snapshot, we partition nodes into different groups (\ie, PP Community and AP Community) and combine PP Communities and AP Communities to gain the initial SC Communities in Section \ref{initializing community}. Subsequently, we classify the dynamic changes into several simple actions, including the adding or removing nodes or edges, and handle them respectively using local information, \ie, a dynamic tracking method in Section \ref{tracking method}.


\subsection{Preliminaries of FOCS}\label{FOCS}

FOCS \cite{overlap} is a community detection algorithm for static and overlapped networks. It is fit for the physical proximity community detection. Comparing with other static detection algorithms (\eg some MODULARITY $Q$-based methods \cite{newman2004finding,newman2004analysis,clauset2004finding,guimera2005functional,blondel2008fast,rosvall2008maps,leicht2008community}, CFinder \cite{palla2005uncovering}, COPRA \cite{gregory2010finding}), it has the following advantages:
\begin{itemize}
   \item It does not have the problems of resolution limit and extreme degeneracy brought about by the MODULARITY $Q$-based method \cite{khadivi2011network}.
   \item It does not need the prior knowledge about the overlapped communities.
   \item It has a good effect on the detection precision.
\end{itemize}

 Considering the MSNs, although FOCS cannot deal with the AP hybrid infrastructure for space-crossing community detection, it is a good choice for our basic physical proximity community detection. Note that, at the technical level, other detection methods also can be used in finding physical proximity community in Section \ref{initializing community}.


\subsection{Initializing Community Structure} \label{initializing community}

\begin{figure} [t]
\begin{center}
\begin{tabular}{cc}
\includegraphics[width=0.5\columnwidth]{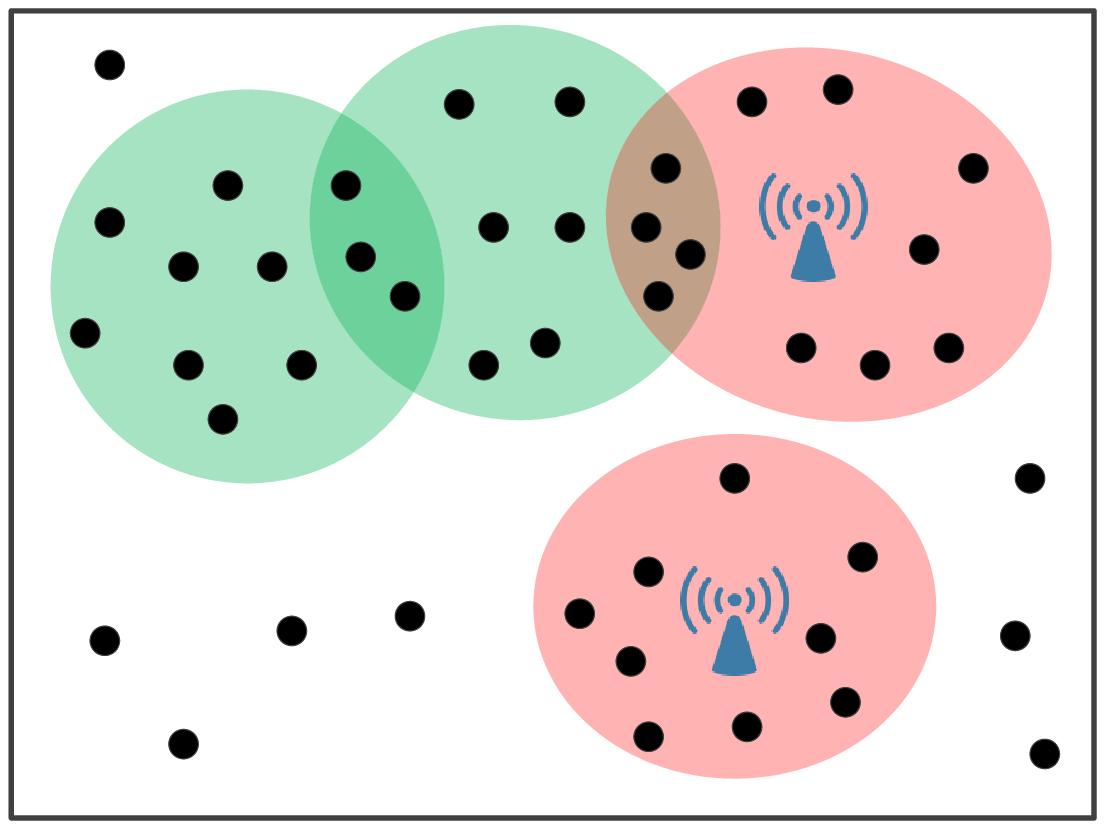}&
\includegraphics[width=0.5\columnwidth]{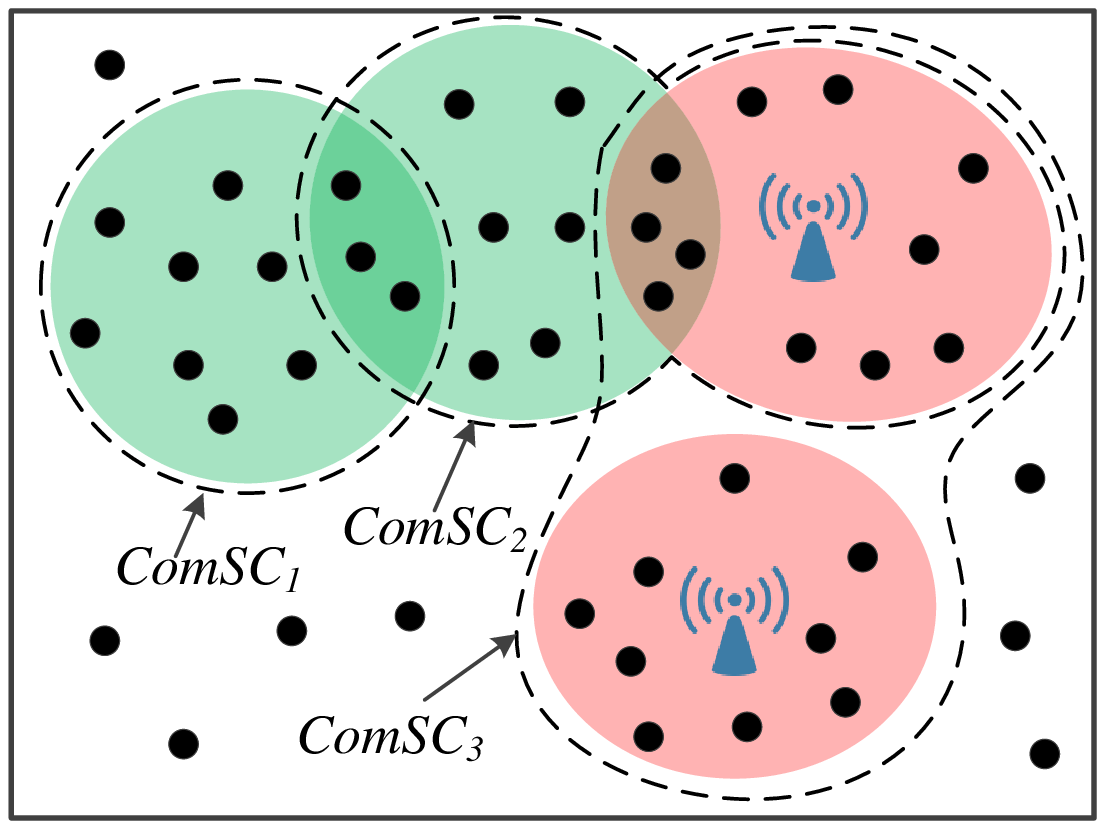} \\
(a) & (b)
\end{tabular}
\vspace {-0.1in}
\caption{Fig(a) shows the results of Step 1-Step 3. Fig(b) shows Step 4. The green dashed areas cover the PP Communities; the red dashed areas cover the AP Communities; the black dash line circles the final SC Communities (In the following figures, we continue to use the same color and line notations). In Fig(b), a PP Community and an AP Community combine into a SC Community $ComSC_{2}$; two AP Communities combine into a SC Community $ComSC_{3}$.} \label{fig2}
  \end{center}
\end{figure}

In the initializing phase, we first need to detect two kinds of communities (\ie, PP Community and AP Community), described in Step 1-Step 3. Then, we require to combine the current communities to gain the initial space-crossing communities, described in Step 4. The combination criteria  are presented in Section \ref{combination}. Fig.\ref{fig2} gives an intuitional presentation of the initializing phase.

1. Circle the area of each AP controlled to get AP Communities. Denote $\emph{i}th$ AP Community at time $t$ as $ComAP_{i}^{t}$.

2. For mobile users, use FOCS algorithm to find PP Communities. Denote $\emph{i}th$ PP Community at time $t$ as $ComPP_{i}^{t}$.

3. Because users can communicate with each other directly through an AP, for each PP Community, if the nodes in a PP Community all have AP marks (including the same AP marks and the different AP marks), we will integrate the PP Community into AP Community(ies), \ie, there do not exist any PP Communities in an AP Community.

4. Combine the overlapped PP Communities and AP Communities according to \textbf{Criterion $\mathcal{S}^{a}$}.  Combine the neighboring AP Communities according to \textbf{Criterion $\mathcal{S}^{b}$}. Then, the current communities are the final Space-Crossing Community (SC Community). Denote $\emph{i}th$ SC Community at time $t$ as $ComSC_{i}^{t}$.

\subsection{Combination Criterion} \label{combination}

\emph{\textbf{Criterion $\mathcal{S}^{a}$}}: If the overlapped/shared substructure of the PP and AP Communities exceeds a threshold value $\alpha$, we will combine them. Threshold $\alpha$ is gained according to  data forwarding experiments in Section \ref{Evaluation}. The shared substructure is measured by a sum of two parts, one is the ratio of the overlapped intra edges to the minimum number of intra edges between two overlapped communities, the other is the ratio of the overlapped nodes to the minimum number of nodes between two overlapped communities.

\emph{\textbf{Criterion $\mathcal{S}^{b}$}}: Considering  the work load (centralized bottleneck) and the communication capability of an AP, even if all APs are connected in a complete graph, APs may not support the multi-hop communication. Therefore, we combine the neighboring/one hop AP Communities  in a clockwise or anticlockwise direction alone the edges between APs (\ie, AP circle) and execute this only once. Finally, we get one community for each pair of APs. This method is one of the practicable solutions. In future, we can make an adaptive scheme according to the current network bandwidth and other factors.

\subsection{Dynamic Tracking Method} \label{tracking method}

After constructing the initial SC Communities, with the passage of time, mobile users will join in or withdraw from the social network and the strength of social relationships will change. Here, with respect to the entire social network, we classify the dynamic network changes into four simple kinds: adding a node, removing a node, adding an edge and removing an edge. We design the following four tracking methods to deal with them respectively.


\subsubsection{Adding A Node}

\begin{figure} [t]
\begin{center}
\begin{tabular}{cc}
\includegraphics[width=0.45\columnwidth]{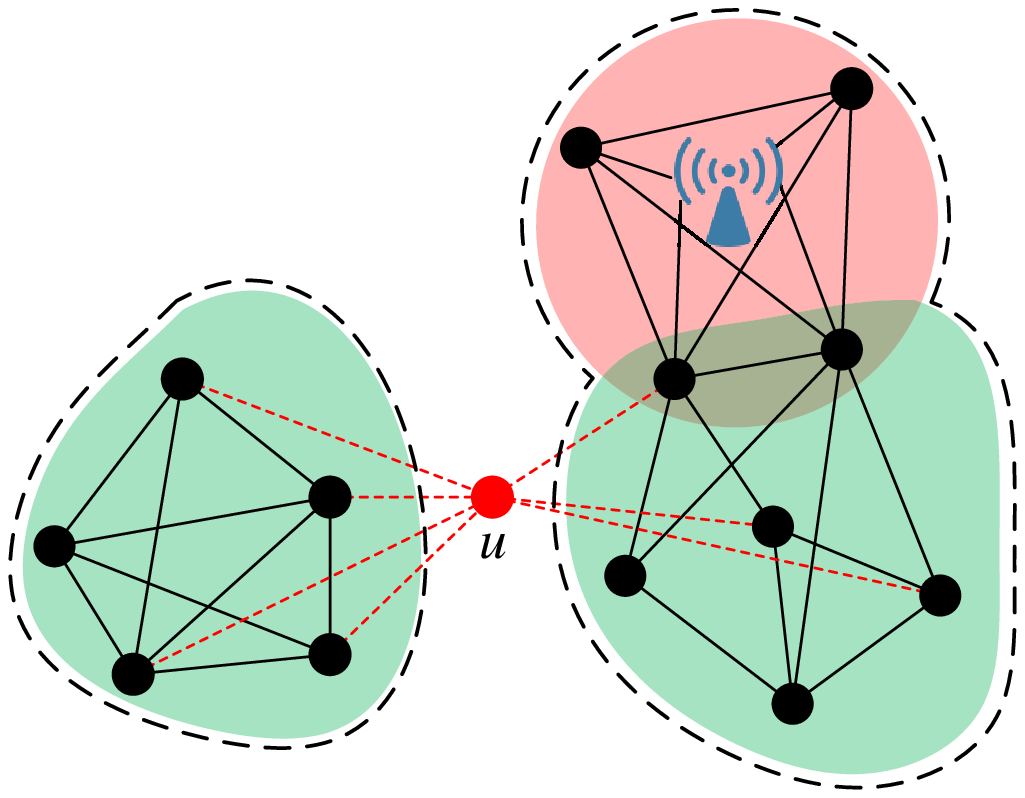}&
\includegraphics[width=0.45\columnwidth]{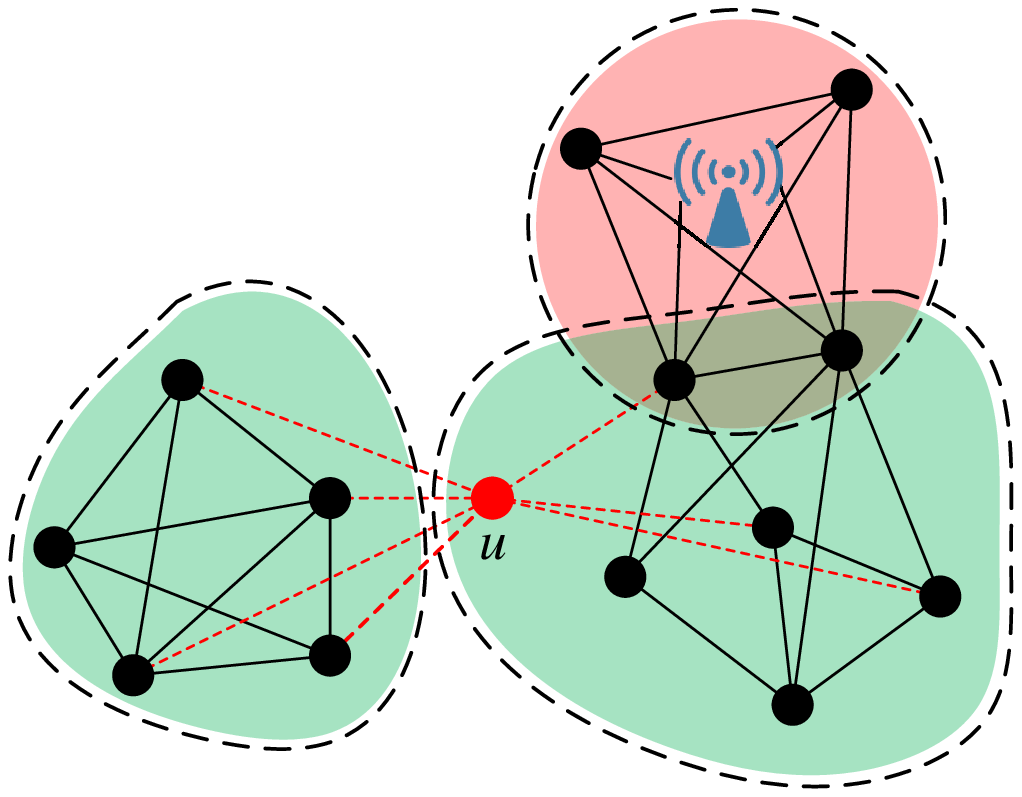} \\
(a) & (b)\\
\includegraphics[width=0.45\columnwidth]{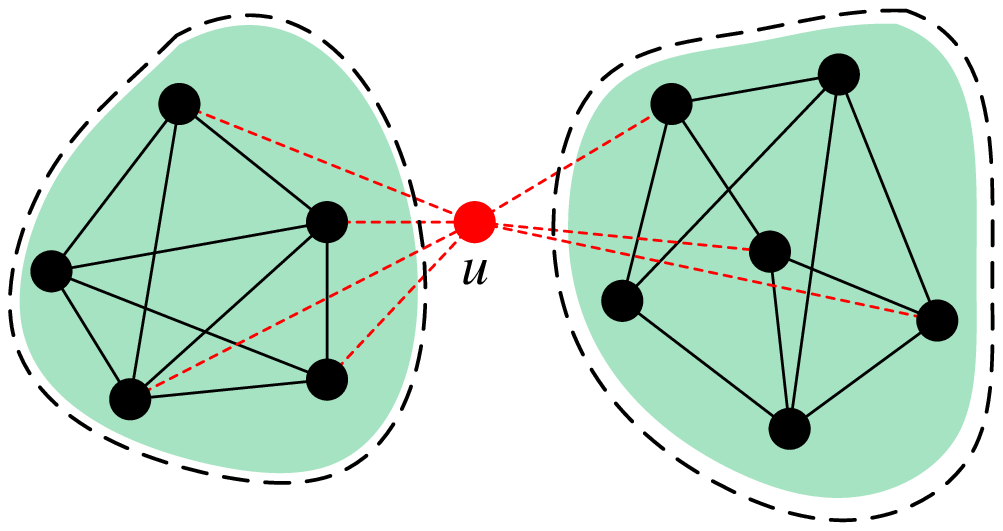}&
\includegraphics[width=0.45\columnwidth]{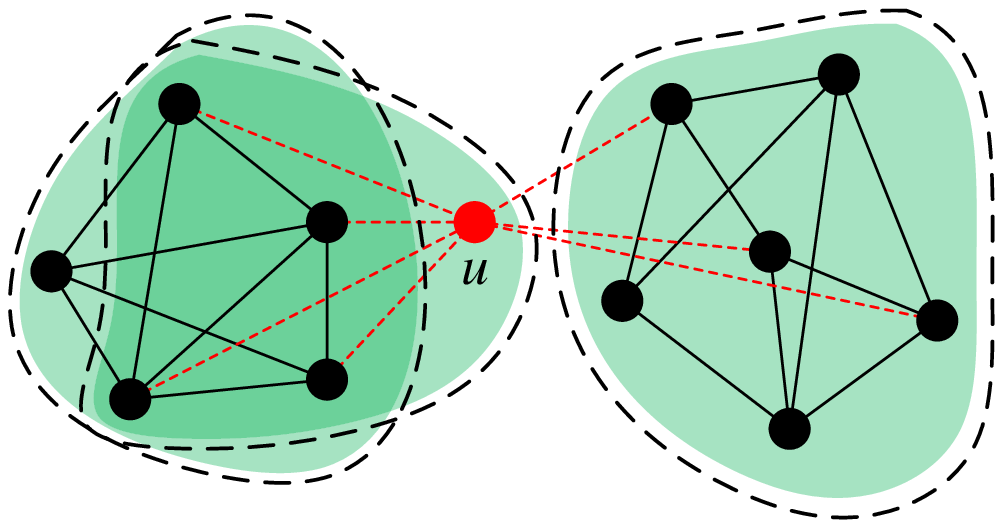} \\
(c) & (d)\\
\includegraphics[width=0.45\columnwidth]{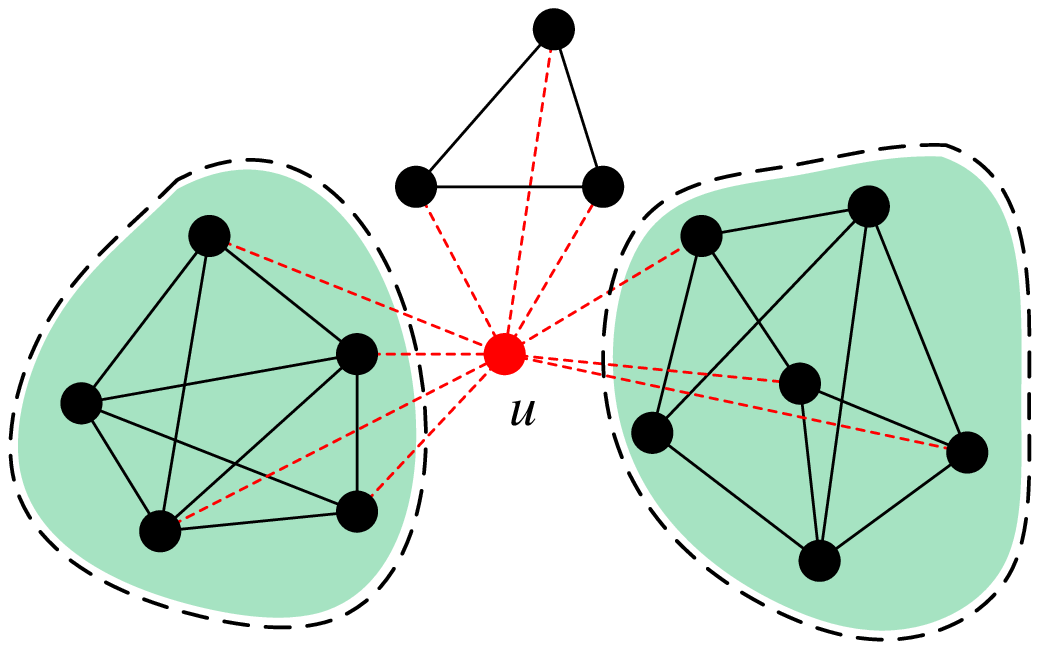}&
\includegraphics[width=0.45\columnwidth]{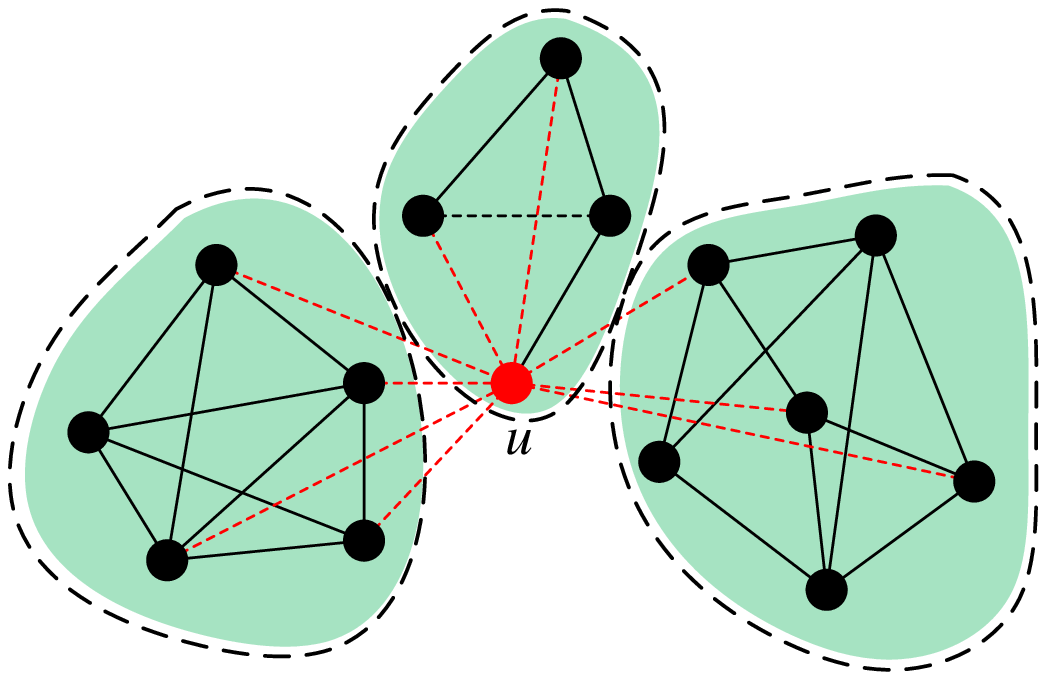} \\
(e) & (f)
\end{tabular}
\vspace {-0.1in}
\caption{Fig(a) and Fig(b) show that the adding node $u$, with associated links (red lash lines), joins in its adjacent right PP Community and the number of SC Communities changes from two to three. Fig(c) and Fig(d) show that, node $u$ and its neighbors forms a new PP Community. Fig(e) and Fig(f) show that node $u$ shapes a new PP Community with the solitary nodes.} \label{fig3}
\end{center}
\vspace{-0.1in}
\end{figure}

%
\begin{itemize}

\item According to \emph{FOCS Community Criterion}, we judge:

  (a) whether the adding node $u$ joins to its adjacent PP Communities;

  (b) whether the intersection of node $u$'s neighbors and adjacent PP Communities forms a new PP Community;

  (c) whether node $u$ shapes a new PP Community with each connected solitary node.

\item If the adding node $u$ is in an AP Community, we add an edge to the AP and add edges with other members belonging to this AP Community to form a complete graph.

\item Find all AP and PP Communities  which are overlapped with the new PP Communities the adding node $u$ belonging  to. We combine these PP Communities with the new PP Communities using \emph{FOCS Combining Criterion} and combine these AP Communities with the new PP Communities using combining criterion $\mathcal{S}^{a}$ to form final new CS Communities. We denote this combining criterion in dynamic tracking phase as \emph{\textbf{ Criterion $\mathcal{S}^{c}$}}.
    Fig.\ref{fig3} is a schematic diagram of \emph{Adding A Node}.

\end{itemize}
%

\subsubsection{Removing A Node}

\begin{figure} [t]
\begin{center}
\begin{tabular}{cc}
\includegraphics[width=0.45\columnwidth]{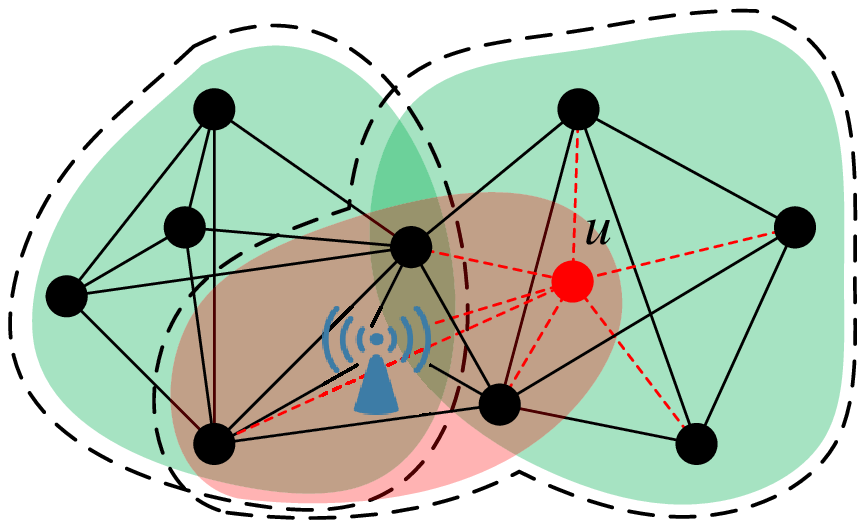}&
\includegraphics[width=0.45\columnwidth]{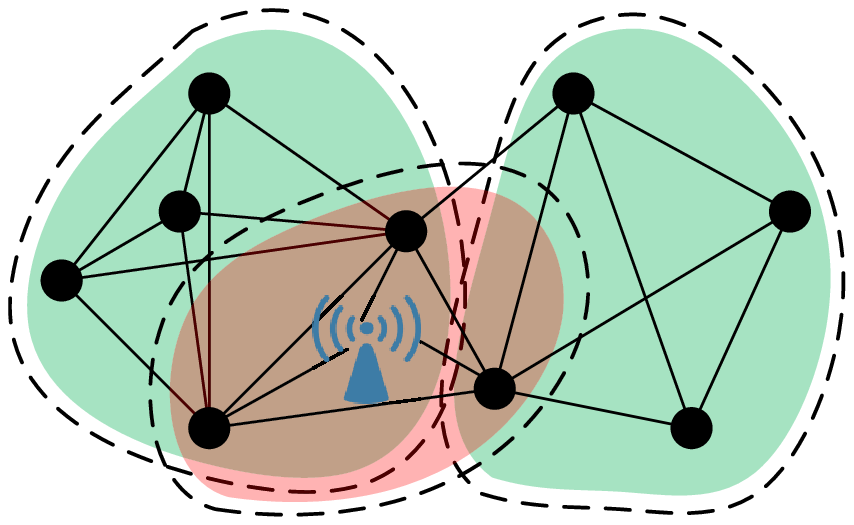} \\
(a) & (b)
\end{tabular}
\vspace {-0.1in}
\caption{Fig(a) shows that the removing node $u$ is in the right PP Community with associated links (red lash lines). After removing node $u$,  Fig(b) shows that the left PP Community maintains, however, the right PP Community changes and the number of SC Communities changes from two to three.} \label{fig4}
\end{center}
\vspace{-0.1in}
\end{figure}

\begin{itemize}

\item Remove the node and its corresponding links from the current network. If the removing node $u$ is in a PP Community, according to \emph{FOCS Community Criterion}, we judge:

 (a) whether the remaining structure can maintain the original PP Community;

 (b) whether the remaining structure forms new PP Communities.

\item Find all AP and PP Communities which are overlapped with the new PP Communities. According to combining criterion $\mathcal{S}^{c}$, we obtain new SC Communities. Fig. \ref{fig4} is a schematic diagram of \emph{Removing A Node}.

\end{itemize}


\subsubsection{Adding An Edge}
%

\begin{figure} [t]
\begin{center}
\begin{tabular}{cc}
\includegraphics[width=0.45\columnwidth]{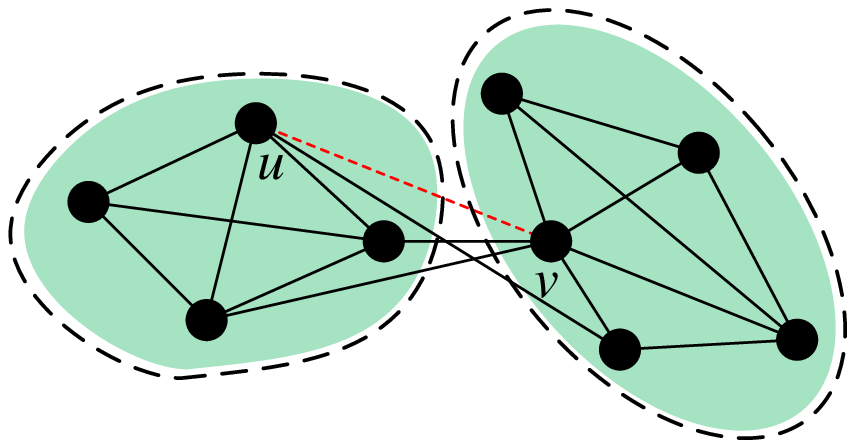}&
\includegraphics[width=0.45\columnwidth]{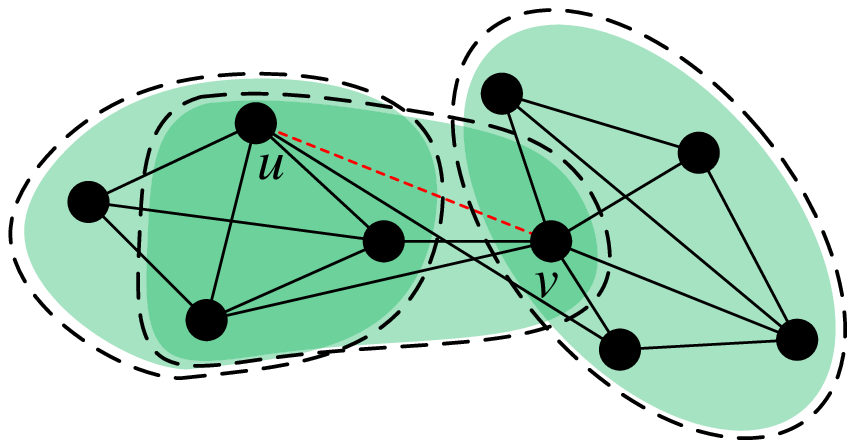} \\
(a) & (b)\\
\includegraphics[width=0.45\columnwidth]{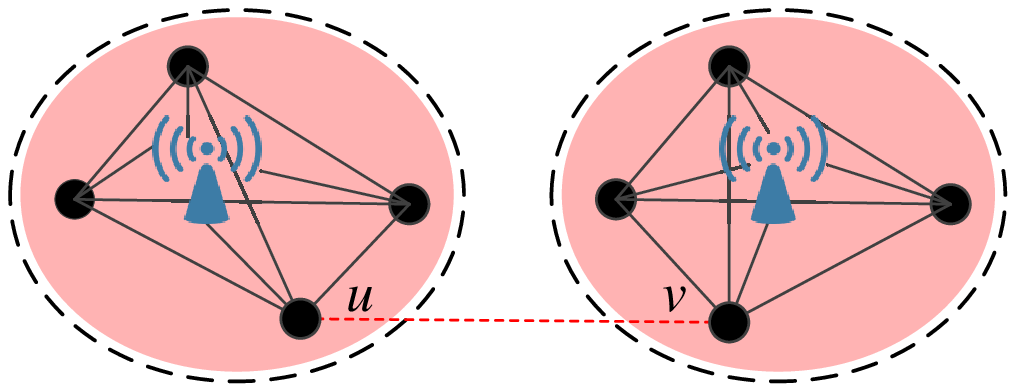}&
\includegraphics[width=0.45\columnwidth]{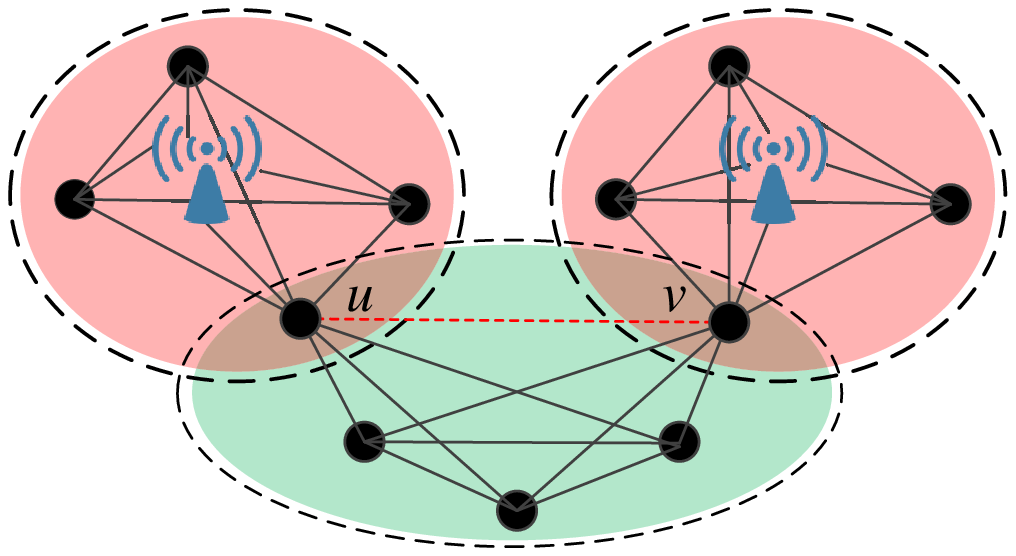} \\
(c) & (d)\\
\includegraphics[width=0.45\columnwidth]{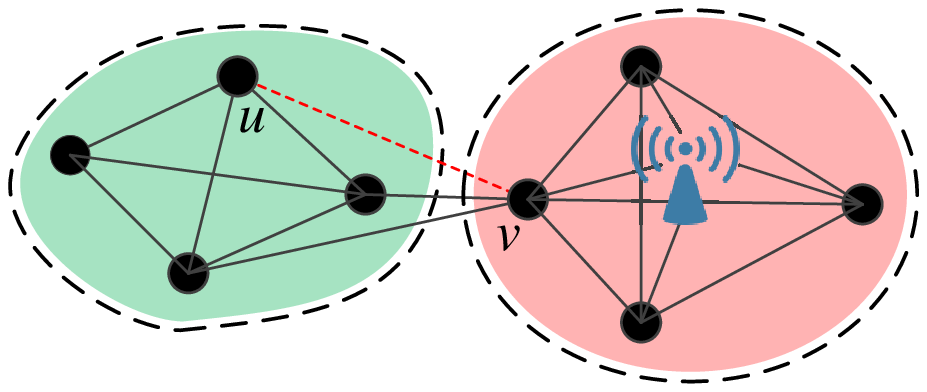}&
\includegraphics[width=0.45\columnwidth]{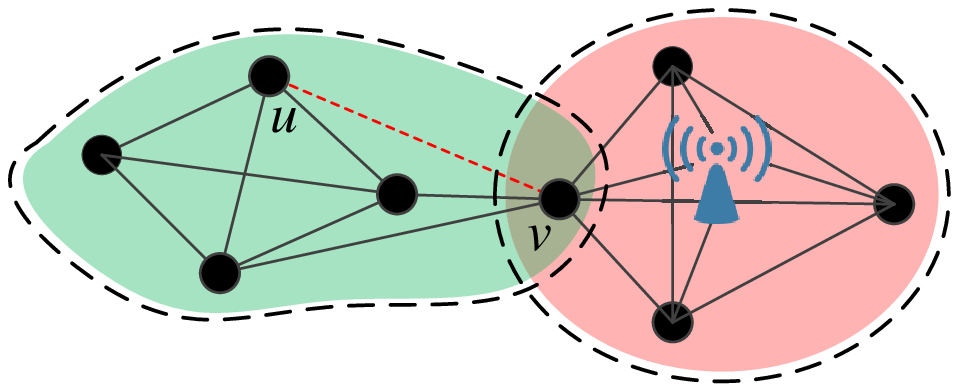} \\
(e) & (f)
\end{tabular}
\vspace {-0.1in}
\caption{Fig(a) and Fig(b) show that, if two endpoints of the adding edge $(u,v)$ are in the different PP Communities, $(u,v)$ forms a new PP Community and the number of SC Communities changes from two to three. Fig(c) and Fig(d) show that, if two endpoints of the adding edge $(u,v)$ are in the different AP Communities, $(u,v)$ forms a new PP Community. Fig(e) and Fig(f) show that, if one endpoint $u$ of the adding edge is in a PP Community, the other endpoint $v$ is in an AP Community, the node $v$ joins in the PP Community.} \label{fig5}
\end{center}
\vspace{-0.1in}
\end{figure}

\begin{itemize}
 \item If two endpoints of the adding edge $(u,v)$ are in the different PP Communities, according to \emph{FOCS Community Criterion}, we judge:

     (a) whether the edge can form a new PP Community;

     (b) whether the node $u$ or $v$ will join in the PP Community of the opposite side.

 \item  If two endpoints of the adding edge $(u,v)$ are in the different AP Communities, according to \emph{FOCS Community Criterion}, we judge:

     (a) whether the edge can form a new PP Community;

 \item  If one endpoint $u$ of the adding edge is in a PP Community, the other endpoint $v$ is in an AP Community, according to \emph{FOCS Community Criterion}, we judge:

     (a) whether the adding edge $(u,v)$ can form a new PP Community.

     (b) whether node $v$ will join in the PP Community which node $u$ belongs to.

\item  Find all AP and PP Communities  which are overlapped with the new PP Community the adding edge $(u,v)$ belonging to. According to combining criterion $\mathcal{S}^{c}$, we obtain new SC Communities. Fig. \ref{fig5} is a schematic diagram of \emph{Adding An Edge}.
\end{itemize}



\subsubsection{Removing An Edge}

\begin{figure} [t]
\begin{center}
\begin{tabular}{cc}
\includegraphics[width=0.38\columnwidth]{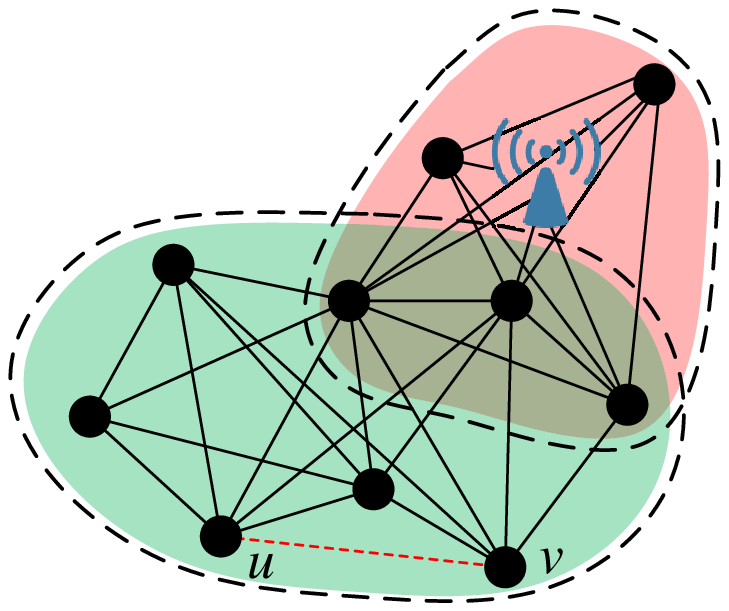}&
\includegraphics[width=0.38\columnwidth]{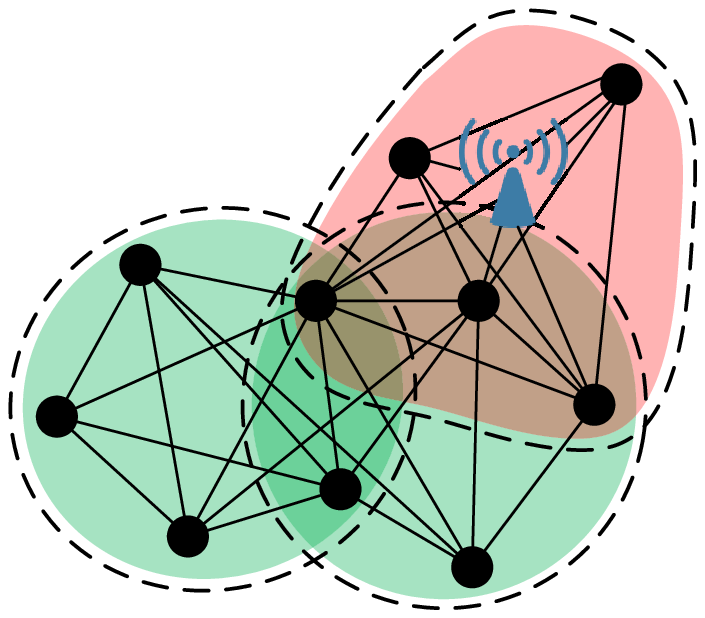} \\
(a) & (b)
\end{tabular}
\vspace {-0.1in}
\caption{Fig(a) shows that the removing edge $(u,v)$ is in a PP Community. After removing the edge,  Fig(b) shows that the PP Community breaks into two PP Communities and the number of SC Communities changes from two to three.} \label{fig6}
\end{center}
\vspace{-0.1in}
\end{figure}

\begin{itemize}

\item Remove the edge $(u,v)$ from the current network. If the two endpoints of the removing edge $(u,v)$ are in the same PP Communities, according to \emph{FOCS Community Criterion}, we judge:

    (a) whether the remaining structure can maintain the original PP Community;

    (b) whether the remaining structure forms new PP Communities.

\item Find all AP and PP Communities which are overlapped with the new PP Communities. According to combining criterion $\mathcal{S}^{c}$, we obtain new SC Communities. Fig. \ref{fig6} is a schematic diagram of \emph{Removing An Edge}.
\end{itemize}

\emph{Remark}: In this paper, APs not only are used to support a hybrid underlying network and bring about the concept of space-crossing community,  but also can undertake some work in community detection and data forwarding. Therefore, we give two hybrid schemes to implement our space-crossing  community detection algorithm.

\emph{ Scheme \uppercase\expandafter{\romannumeral1}}: One-layer AP hybrid infrastructure, \ie, APs are deployed in the network area, not covering the entire area. In AP Communities, the AP centralized way is applied. An AP knows its members and the relationships among them. In non-AP Communities, the distributed or non-organized  way is applied. Nodes undertake a large portion of work. A node have perfect knowledge of its neighbors and some local approximation knowledge captured by its neighbors.  Some required information is transferred through node to node. For example, a  powerful node can be assigned to deal with some local changes.

\emph{Scheme \uppercase\expandafter{\romannumeral2}}: Multi-layer AP hybrid infrastructure. For example, there are two layers and the top layer AP is used as a supervisor to control all nodes in the network. This supervisor is not unreasonable to be set because some global work execute only once in the algorithm, such as the initializing community structure phase.

Two schemes mentioned above are hybrid. They can make the network load in balance, i.e., the work is shared among nodes and APs.

%
\section{SAAS Data Forwarding Scheme}\label{data forwarding}

In this section, based on  Space-Crossing Community Detection results, we design a SAAS (Similarity Attraction and AP Spreading) data forwarding scheme to validate the positive role of the space-crossing community.

\subsection{Pearson Social Similarity} \label{Similarity Attraction}

\begin{defn}[Local Activity]\label{Activity}

Let $a_{u,i}^{t}$  denote the local activity
of node $u$ in a space-crossing community $ComSC_{i}^{t}$ at time $t$. Then,
$$a_{u,i}^{t}=\frac{\sum_{(u,v) \in ComSC_{i}^{t}} w_{uv}^{t}}{\sum_{(v^{'},v^{''})\in ComSC_{i}^{t}}w_{v^{'}v^{''}}^{t}}, 1\leq i \leq k, v\neq u$$
where $v^{'}$ and $v^{''}$ are any two nodes in $ComSC_{i}^{t}$; $w_{uv}^{t}$ has been defined in Section \ref{Contact Aggregation}; $k$ represents the number of SC Communities; the numerator represents the sum of the encounter ratio between node $u$ and other nodes in community $ComSC_{i}^{t}$ and the denominator represents the sum of the encounter ratio between any two nodes in community $ComSC_{i}^{t}$.
\end{defn}

Node local activity can represent the importance of a node in a certain community. A larger local activity means that the node has more
interactions with other members in the community.
 In data forwarding, local activity is important because if the message is given to a node having low local activity, it will bring about a low efficiency in terms of delivery ratio.

There exist some other methods and concepts which may
be confused with our local activity. We give the detailed
explanations so as to distinguish them. In
Simbet \cite{daly2007social} and BUBBLE RAP \cite{hui2008bubble}, they use betweenness
centrality in data forwarding. Betweenness measures the
extent to which a node lies on the shortest paths linking
other nodes. A node with a high betweenness centrality has
a capacity of facilitating interactions between the nodes that
it links. However, not only global but also local centrality
are only fit for unweighted graphs. In unweighted graphs,
there will be an edge if there exists a contact between two
nodes. But in reality, the contact probability may be too low
to be utilized in data forwarding, i.e., betweenness centrality
can not reflect the encounter probability. To some extent,
local centrality and local activity both can represent the
importance of a node in its communities, but, they are not
the same concept. The former is only with node degrees,
the latter is a statistics of encounter probability in a node
certain communities.

\begin{defn}[Activity Vector]\label{Activity Vector}
We define an activity vector $A_{t}(u)=(a_{u,1}^{t},a_{u,2}^{t},...,a_{u,i}^{t},...,a_{u,k}^{t})$ for each node $u$ at time $t$, where $a_{u,i}^{t}$ denotes the local activity of node $u$ in space-crossing community $ComSC_{i}^{t}$ at time $t$. The value of $k$ represents the number of communities after applying the space-crossing community detection method.
\end{defn}

There are some social similarity measurements which are often used in previous studies, such as cosine angular distance\cite{growingnetworks}, Hamming feature distance\cite{wu2012social}, the number of common communities (interests groups)\cite{overlap}. However, the \emph{distance-based} methods cannot give a  meaningful explanation in real social networks. The \emph{common interests-based} method has a problem that if we choose a node having more  common communities with the destination as a relay node, the chosen node may be one with low local activity in its community. In this paper, we introduce Pearson Correlation Coefficient method to define social similarity.

\begin{defn}[Pearson Social Similarity]\label{Social Similarity}
Given two activity vectors $A_{t}(u)=(a_{u,1}^{t},a_{u,2}^{t},...,a_{u,i}^{t},...,a_{u,k}^{t})$ of node $u$ and $A_{t}(w)=(a_{w,1}^{t},a_{w,2}^{t},...,a_{w,i}^{t},...,a_{w,k}^{t})$ of node $w$, we define the social similarity between $u$ and $w$ at time $t$ as $SS_{t}(u,w)$, having
\begin{center}
$SS_{t}(u,w)=\frac{E(A_{t}(u)A_{t}(w))-E(A_{t}(u))E(A_{t}(w))}{\sqrt{E(A_{t}(u)^{2})-E^{2}(A_{t}(u))}\sqrt{E(A_{t}(w)^{2})-E^{2}(A_{t}(w))}}$.
\end{center}
\end{defn}

Pearson correlation coefficient reflects the degree of linear dependence between two vectors. Local activity reflects the importance of a node in a certain space-crossing community.
\emph{Pearson social similarity} is a combination calculation of local activity and pearson correlation coefficient.
Assuming that node $w$ is the destination node. There exists a session from node $u$ to node $w$. The candidate  relay  is node $v$. If the \emph{pearson social similarity} $SS_{t}(v,w)$ is larger than $SS_{t}(u,w)$, then, we can say the degree of linear dependence between $v$ and $w$ is larger than $u$ and $w$. In each vector component, node $v$ and $w$ are more anastomotic. Reflected in social networks, they are more similar in social aspects, \ie, the interests groups and the local activity of node $v$ are proportionate to  node $w$. Intuitively, if the destination node is in $ComSC_{1}^{t}$ and $ComSC_{2}^{t}$ two interests groups at time $t$, and the node local activity in $ComSC_{1}^{t}$ is larger than in $ComSC_{2}^{t}$. Then, the node that has the same characteristic with the destination is more appropriate as a relay than that has not. Thus, a larger \emph{pearson social similarity} can guarantee the relay node having high chance to forward data successfully.

\subsection{SAAS Algorithm} \label{Similarity Attraction}

\begin{figure} [t]
\begin{center}
\includegraphics[width=0.6\columnwidth]{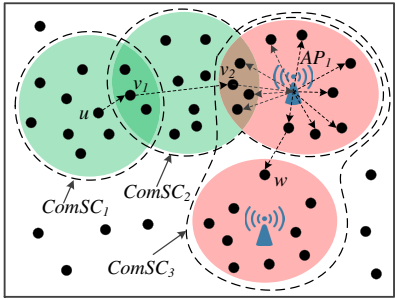}
\vspace {-0.1in}
\caption{A session from node $u$ to $w$. The black dotted arrow line means a node mobile and forwarding path. Similarity attraction is taken place in the course of node $u$ to $v_{1}$ and node $v_{1}$ to $v_{2}$. When node $v_{2}$ receives the message, through Access Point $AP_{1}$, the message copies spread in the AP Community. Finally, one of the copies is transferred to the destination node $w$.} \label{fig7}
\end{center}
\vspace{-0.1in}
\end{figure}

In this section, we provide the details of SAAS (Similarity Attraction and AP Spreading) algorithm.
It includes two phases: \emph{Similarity Attraction Phase} and \emph{AP Spreading Phase}, as illustrated in Fig. \ref{fig7}.

\subsubsection{Similarity Attraction Phase}

In this phase, each node has its activity vector. When the message holder meets another node, they will calculate their \emph{pearson social similarity} with the destination respectively. The message holder tries to send the message
to a node which has larger \emph{pearson social similarity} than it has and let the node send the message
to the destination consecutively.

\subsubsection{AP Spreading Phase}

In this phase, if the message holder enters into an AP Community $ComAP_{i}^{t}$, through the AP, it will give each mobile node within $ComAP_{i}^{t}$  a message copy and let the nodes jointly send copies to the destination. Before reaching the destination, the more nodes the message copies are sent to, the smaller delay  our SAAS algorithm
will have. Note that, in this phase, the similarity attraction does not work due to spreading copies. That is to say, in AP Communities, if the encountered  node has higher \emph{pearson social similarity} with the destination than the current message holder, the operation of sending the message will not occur between the holder and the encounter.

We present  SAAS algorithm, as shown in Algorithm \ref{Algorithm 1}. We do not distinguish the two phases in sequence. This is because the message
exchange in SAAS is compatible with each phase.

\emph{Remark 1}: SAAS  is a distributed algorithm, in which each node
only requires to calculate its \emph{pearson social similarity} with the destination and decides whether to send the message to the encountered node or spread message copies in an AP Community.

\emph{Remark 2}: To keep the knowledge of each node synchronous, we update the community list periodically. This time interval is set according to the WIFI scanning time in the dataset.

\emph{Remark 3}: With regard to dissemination among APs, we assume that an Ap does not talk to another AP, it is only their configuration (circle) for analysis of community detection.

\begin{algorithm}[h]
\caption{SAAS: a session from node $u$ to $w$ at time $t$}
\label{Algorithm 1}
\begin{algorithmic}[1]
\algsetup{linenosize=\footnotesize}
\footnotesize

\STATE{if message holder $u$ is in an AP Community $ComAP_{i}^{t}$}
\STATE{~~if nodes within $ComAP_{i}^{t}$ have not the message copies}
\STATE{~~~~node $u$ will transmit copies to all nodes within community $ComAP_{i}^{t}$}
\STATE{else}
\STATE{~~if node $u$ encounters another node $v$ without a message copy}
\STATE{~~~~calculate $SS_{t}(u,w)$ and  $SS_{t}(v,w)$}
\STATE{~~~~if $SS_{t}(v,w)>SS_{t}(u,w)$ }
\STATE{~~~~~~node $u$ transmits the message to node $v$}
\STATE{~~~~else}
\STATE{~~~~~~node $u$ maintains the message}
\end{algorithmic}
\end{algorithm}

%




%
%
\section{Performance Evaluation} \label{Evaluation}
In this section, we evaluate the performance of our SAAS data forwarding scheme.
The only parameter  required to be fixed is the combining threshold value $\alpha$ in combining criterion $\mathcal{S}^{a}$. Through setting different values, we work at obtaining an optimal value of $\alpha$ for a high efficient data forwarding. In addition, we evaluate SAAS on two kinds of social datasets. One is dense, the other is sparse. We will test if parameter $\alpha$ is related to the social density of MSNs.


\subsection{Dataset Selection}
\begin{figure*} [t]
\begin{center}
\begin{tabular}{ccc}
\hspace {-0.2in}
\includegraphics[width=0.7\columnwidth]{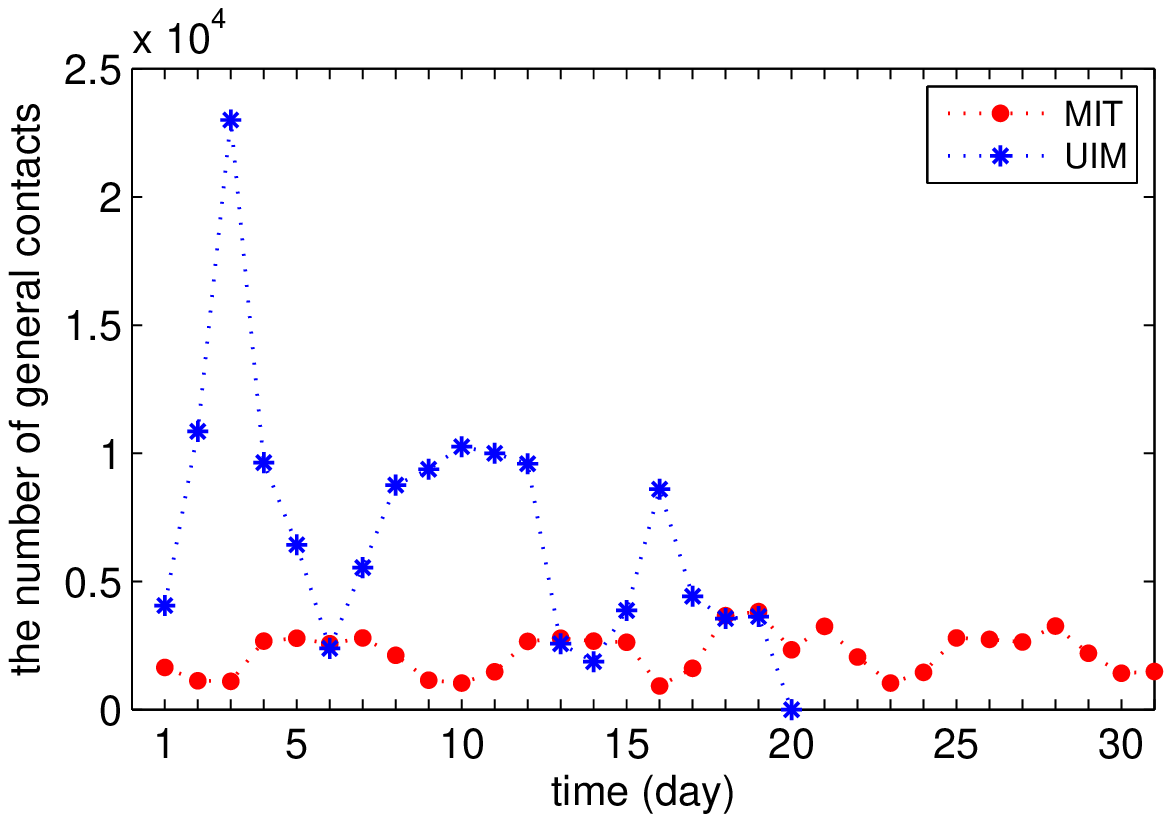}&
\hspace {-0.2in}
\includegraphics[width=0.7\columnwidth]{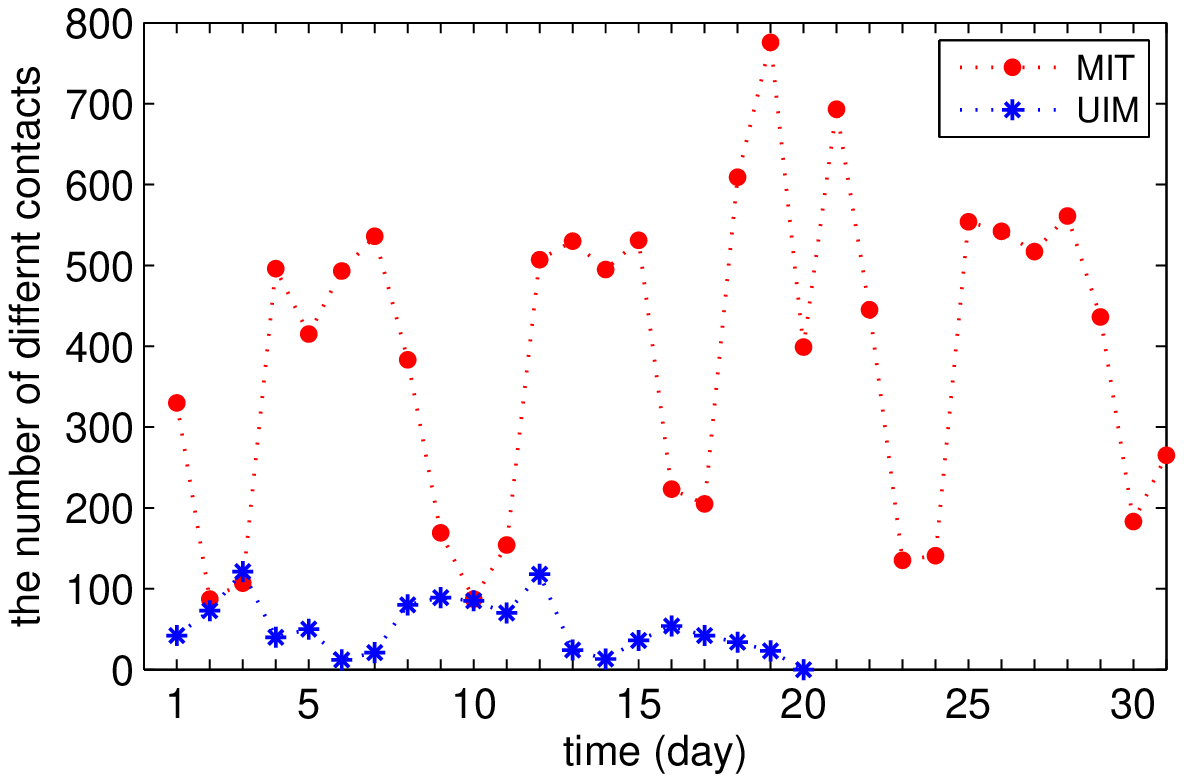}&
\hspace {-0.2in}
\includegraphics[width=0.7\columnwidth]{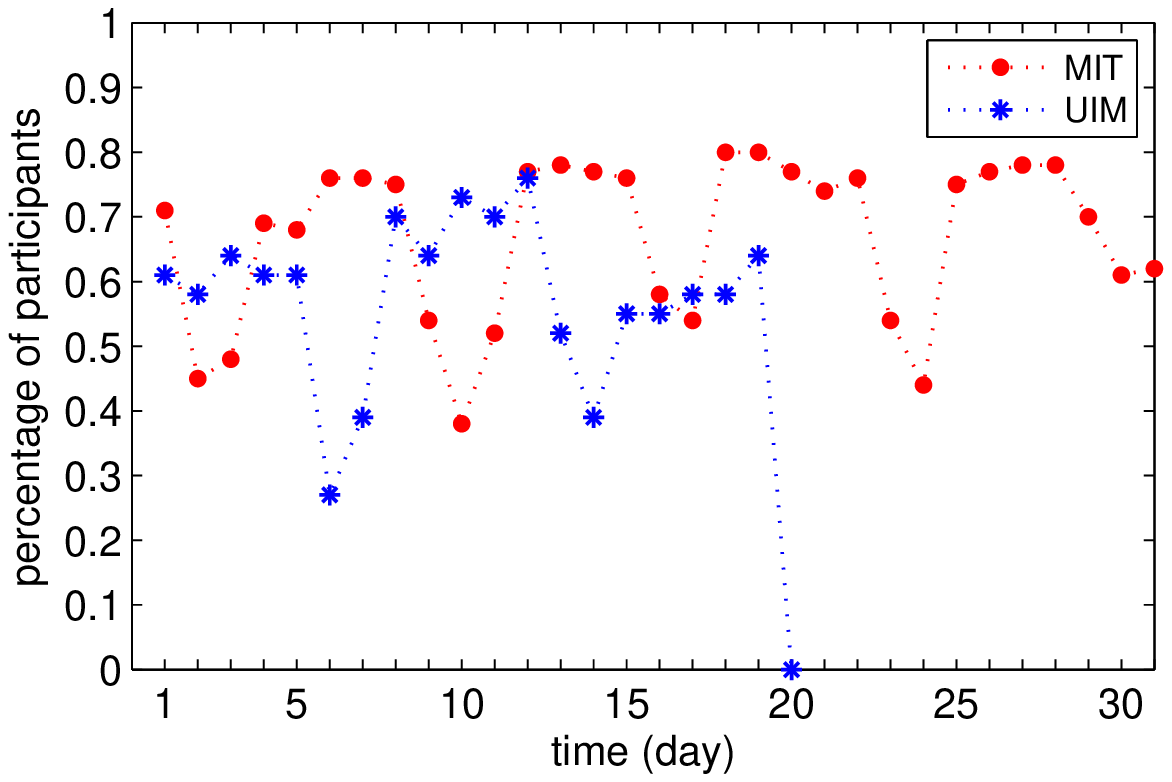}\\
(a) & (b)&(c)
\end{tabular}
\vspace {-0.1in}
\caption{Fig(a), Fig(b) and Fig(c) show the number of contacts, the number of different contacts and  the percentage of participants of MIT and UIM per day respectively.  Note that for UIM, in original dataset,  records of the 20th day are null.}  \label{contact}
\end{center}
\vspace{-0.1in}
\end{figure*}

The evaluation of SAAS is based on MIT Reality Mining \cite{uiuc-uim-2012-01-24} and UIM (University of Illinois Movement) \cite{eagle2009inferring}. From the trace analysis as  illustrated Fig \ref{contact},
MIT contains 2188.71 general contacts, 387.23 different contacts\footnote{The general contact means the sum of all contacts among nodes in a period of time. The different contact means that, if there is more than one contact between two nodes, we will only count the number of contacts once.}and 67.19 percent of  participants per day on average. While, UIM contains 6920.9 general contacts, 51.35 different contacts and  55.25 percent of participants per day on average.  We can see, the participation degree of MIT and UIM is almost the same, however, the number of general contacts and the number of different contacts  for MIT and UIM are opposite starkly. On average, there are 3.46 and 1.57 different contacts \emph{per node} for MIT and UIM respectively. It means that, through contact aggregation (\ie, the weight filtering method in Section \ref{Contact Aggregation}),  the number of edges attached a node in MIT is larger than in UIM. Thus, in our simulation, we say that MIT is a dense social network, while UIM is a sparse one. Note that the concept of  dense/sparse network is just relative.

In MIT Reality Mining, $97$ Nokia $6600$ mobile phones were carried by users over the course of $9$ months in MIT campus and its surroundings. In the long-term observation, the dataset records the contacts between mobile users and the contacts between users and visible GSM cell towers.
In UIM, $28$ Google Android phones were carried by users over the course of  $3$ weeks in university of Illinois. It is a dataset that contains MACs of Bluetooth and WiFi access points captured by phone plug-in middleware periodically.

In two datasets mentioned above, for cell-phone calling requirements, the number of APs is large and APs cover almost entire network. In order to model our hybrid underlying network, we only select $15$ APs and $5$ APs  at random for MIT Reality Mining and UIM respectively. That it to say, the controlled areas of APs do not cover entire network.
Then, according to the method of contact aggregation for edges in Section \ref{Network Model}, we construct social graphs using the scanning records in those datasets. For MIT Reality Mining, we choose the weighted
growing window mechanism. For UIM, we choose the weighted
sliding window mechanism.

\subsection{Simulation Setup}
We choose the ONE simulator as our experimental tool \cite{keranen2009one}.
It not only provides various mobile models including some complex mobility scenarios in daily life, but also can incorporate real world traces into the simulator. According to the contacts among Bluetooth devices and the contacts between Bluetooth devices and APs, we extract trace files from MIT Reality Mining and UIM datasets. These discrete contact events can be taken as the inputs of the ONE simulator. The datasets include the start time, end time and communication peers. We set the start time as connection up and the end time as connection down. One example of the trace file extracted from MIT Reality Mining is shown as follows:
\[
  \begin{array}{ccccc}
    0 & CONN & 93 & 96 & up \\
    0 & CONN & 93 & 14 & up \\
    128 & CONN & 85 & 17 & up \\
    129 & CONN & 94 & 29 & up \\
      &   & \cdots &   &   \\
    1169 & CONN & 28 & 5 & down \\
    1169 & CONN & 28 & 17 & down \\
    &   & \cdots &   &
  \end{array}
\]

\vspace{-0.1in}
For all simulations in this work, each node generates $1000$ packets during the simulation time. The packet size is distributed from $50KB$ to $100KB$ uniformly.  In the hybrid underlying network, the interface of users (cell-phone carriers) is assigned to two modes: Bluetooth and WIFI, and the interface of APs is assigned to WIFI. Data transmission speed of Bluetooth is $2Mbps$ and transmission range is in $10m$.
Data transmission speed of WIFI is $5Mbps$ and transmission range is in $100m$.
 The scanning interval of Bluetooth is $5min$ and $1min$ for MIT Reality Mining and UIM respectively. The scanning interval of WIFI is $30min$.
The buffer size of each node is $5MB$. The source and destination pairs are chosen randomly among all nodes. Each simulation is repeated $20$ times with different random seeds. Without losing precision, we set the update interval is $1$. For MIT Reality Mining, we set TTL from $30$min to $1$mon\footnote{MIT Reality Mining Dataset is a long-term observation dataset. Thus, some cumulative social phenomena (local activity, community structure \etc) require a period of time to reveal. We do experiments from date  $2004-10-01$ to date $2004-10-31$, \ie, a large TTL-1 month, instead of several days. A larger TTL(larger than 1 month) also can be done with more simulation time and the overall trend is similar with 1 month.}. For UIM, we set TTL from $30$min to $3$week. In these settings, some come from the MIT and UIM datasets, others are set according to common sense. They are not the determinant parameters in SAAS and the following comparison algorithms.


\subsection{Metrics}
\begin{itemize}
\item Delivery Ratio: the ratio of the number of successfully delivered messages to the total number of created messages.
\item Average Latency: the average messages delay for all the successful sessions. The unit of average latency is second.
\end{itemize}

\subsection{Experiments on SAAS Data Forwarding Scheme}
\subsubsection{Comparisons with Other Forwarding Schemes}
$\\$
In this section, we set parameter $\alpha$=0.2, 0.4, 0.6, 0.8 respectively\footnote{We also can set parameter $\alpha$=0.1, 0.3, 0.5, 0.7, 0.9. Because our goal is testing the density sensitivity of parameter $\alpha$, here we only give the evaluation of approximate even values as representatives.} and compare our SAAS algorithm against BUBBLE RAP \cite{hui2008bubble} and Nguyen's Routing \cite{overlap} (\ie, two social community-based routing algorithms).

In BUBBLE RAP, it provides a hierarchical forwarding strategy. A node first bubbles the message up the hierarchical ranking tree using the global centrality. When the message reaches the community of the destination node, local centrality is used instead of the global centrality. In Nguyen's Routing, a smart community detection algorithm is proposed and applied to data forwarding in mobile networks. A message is forwarded to an encountered node if the node shares more common communities with the destination than the current one.  Note that we select settings or parameters which bring about the best performances for above two contrastive algorithms respectively.

\begin{figure*} [t]
\begin{center}
\begin{tabular}{cc}
\hspace {-0.2in}
\includegraphics[width=0.8\columnwidth]{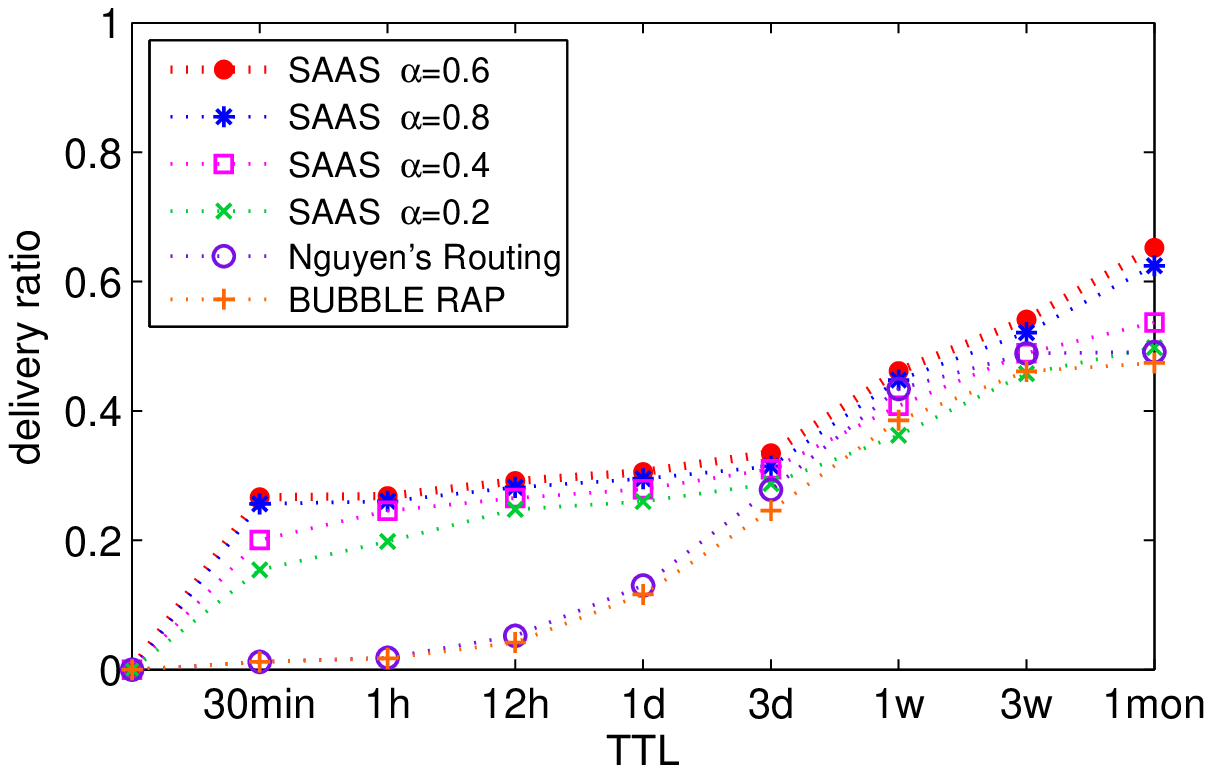} &
\hspace {-0.2in}
\includegraphics[width=0.8\columnwidth]{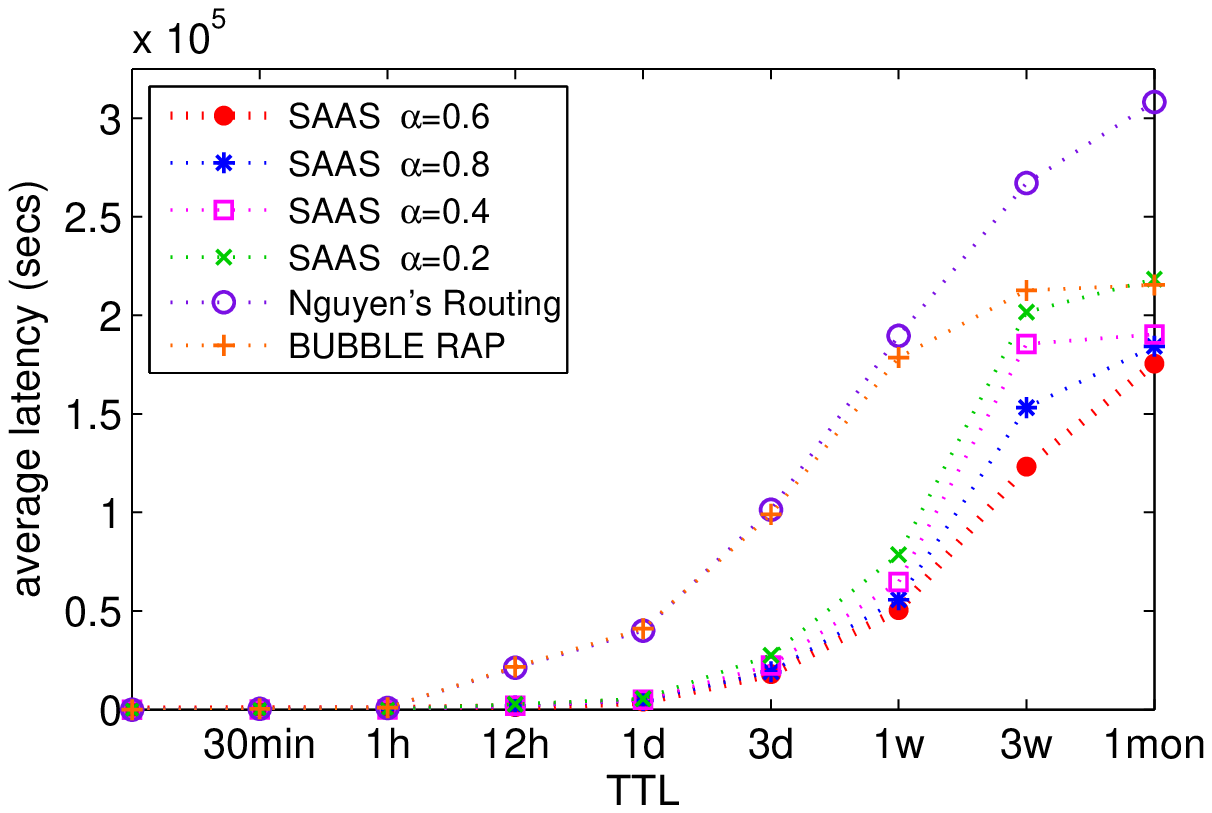}\\
(a) & (b)
\end{tabular}
\caption{Simulation Results on MIT Reality Mining Dataset}\label{exp1}
\end{center}
\vspace{-0.1in}
\end{figure*}
\begin{figure*} [t]
\begin{center}
\begin{tabular}{cc}
\hspace {-0.2in}
\includegraphics[width=0.8\columnwidth]{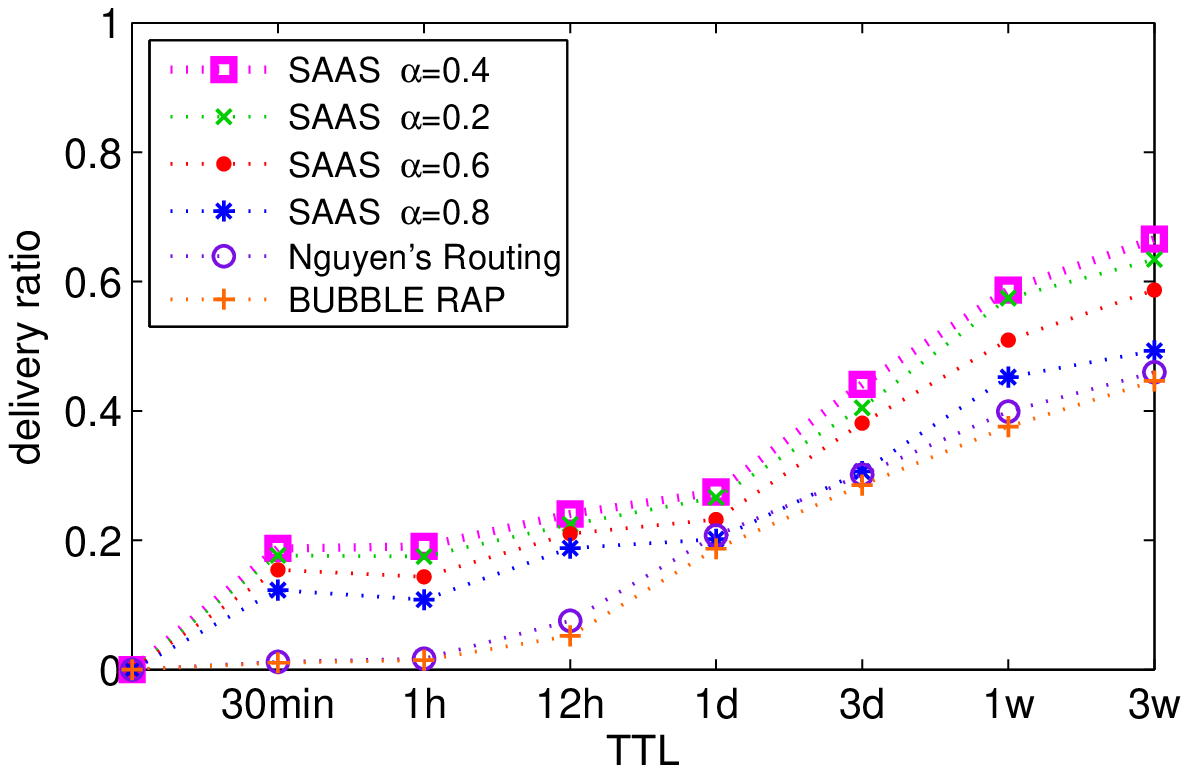}&
\hspace {-0.2in}
\includegraphics[width=0.8\columnwidth]{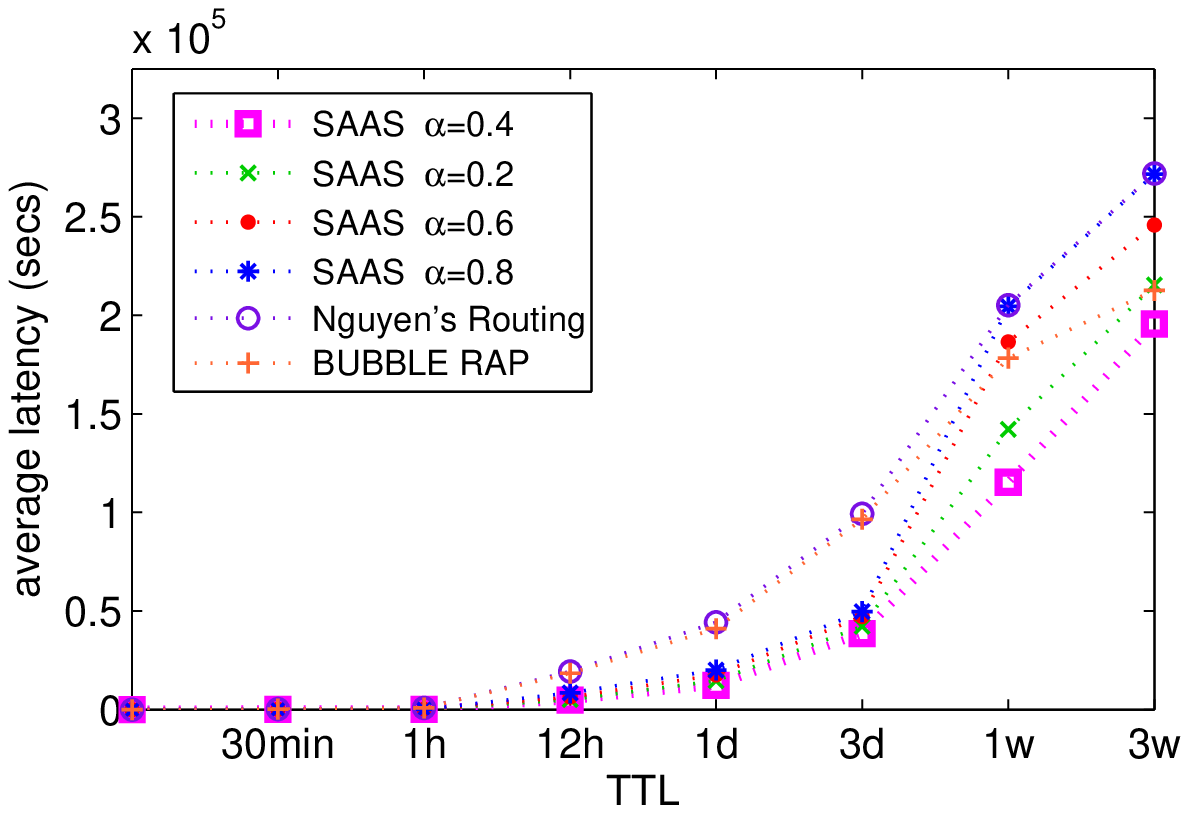}\\
(a) & (b)
\end{tabular}
\caption{Simulation Results on UIM Dataset} \label{exp2}
\end{center}
\vspace{-0.1in}
\end{figure*}

Fig.\ref{exp1} and Fig.\ref{exp2} show the delivery ratio and average latency of our SAAS, BUBBLE RAP and Nguyen's Routing algorithms in MIT Reality Mining and UIM datasets respectively.

 From Fig.\ref{exp1} (a) and Fig.\ref{exp1} (b), we can see,  parameter $\alpha$=0.6 is optimal for MIT Reality Mining (dense social networks). When $\alpha$=0.6, the delivery ratio of SAAS achieves the best among those algorithms and the average latency is lowest. For the same reason,  from Fig.\ref{exp2} (a) and Fig.\ref{exp2} (b), parameter $\alpha$=0.4 is optimal for UIM (sparse social networks).
This is because SAAS tends to combine the overlapped PP and AP Communities to utilize APs  for data forwarding.  For example, if the overlapped PP and AP Communities do not combine (\ie, they belong to different SC Communities), the two activity vectors of two nodes (one is as a message holder in a PP Community, the other is as the encountered node in an AP Community) will differentiate largely in different vector components. Thus, the \emph{pearson social similarity} between the two nodes will be small. The result is the encountered node in the AP Community will be abandoned as a relay. If they combine,  the values of the two nodes' local activity vectors will be similar in components. The \emph{pearson social similarity} between the two nodes will be large which can make APs play roles in data forwarding. But, it does not mean we will combine any overlapped PP and AP Communities. Thus, an optimal parameter $\alpha$ is required for different scenarios.
For the dense network, the probabilities of large numbers of communities and large scale of communities are higher than the sparse one. Thus, the combining parameter $\alpha$ should be large in order to differentiate the different SC Communities. If $\alpha$ is small, there will emerge a large SC Community to make our SAAS invalid.


In MIT Reality Mining dataset, Fig.\ref{exp1} (a) shows that SAAS with $\alpha$=0.6 performs best among those algorithms.
Its delivery ratio is higher than Nguyen's Routing with
63.87 percent, BUBBLE RAP with 77.91 percent
on average. In UIM dataset, Fig.\ref{exp2} (a) shows that SAAS with $\alpha$=0.4 performs best among those algorithms.
Its delivery ratio is higher than Nguyen's Routing with 75.78
percent, BUBBLE RAP with 88.77 percent
 on average.
In terms of delivery ratio, at the initial phase, due to the help of APs, it is obvious that SAAS increases quickly in both MIT and UIM datasets.
 BUBBLE RAP use betweenness as centrality metrics without considering node contact frequency. As long as there exists an edge between two nodes,  the edge will be used in  betweenness calculation. But in real social networks, the edge may only have a trivial effect in data forwarding. Thus, it has a lower delivery ratio than SAAS. Nguyen's Routing tends to send messages
to nodes having many interests with the destination, however, it may deliver them to nodes which have low local activity in their communities (or interests groups). It is the main reason for the low delivery ratio of Nguyen's Routing.
Fig.\ref{exp1} (b) and Fig.\ref{exp2} (b) show that the delays of those algorithms all go up with TTL increasing. SAAS shows the predominant performance among them. Due to the help of SC Communities, some long-distance nodes can communicate through short-path across the geographical space, which lead to the low delay of SAAS.

 \subsubsection{The Role of Space-Crossing Community}
 $\\$
 \begin{figure} [t]
\begin{center}
\begin{tabular}{cc}
\hspace {-0.2in}
\includegraphics[width=0.55\columnwidth]{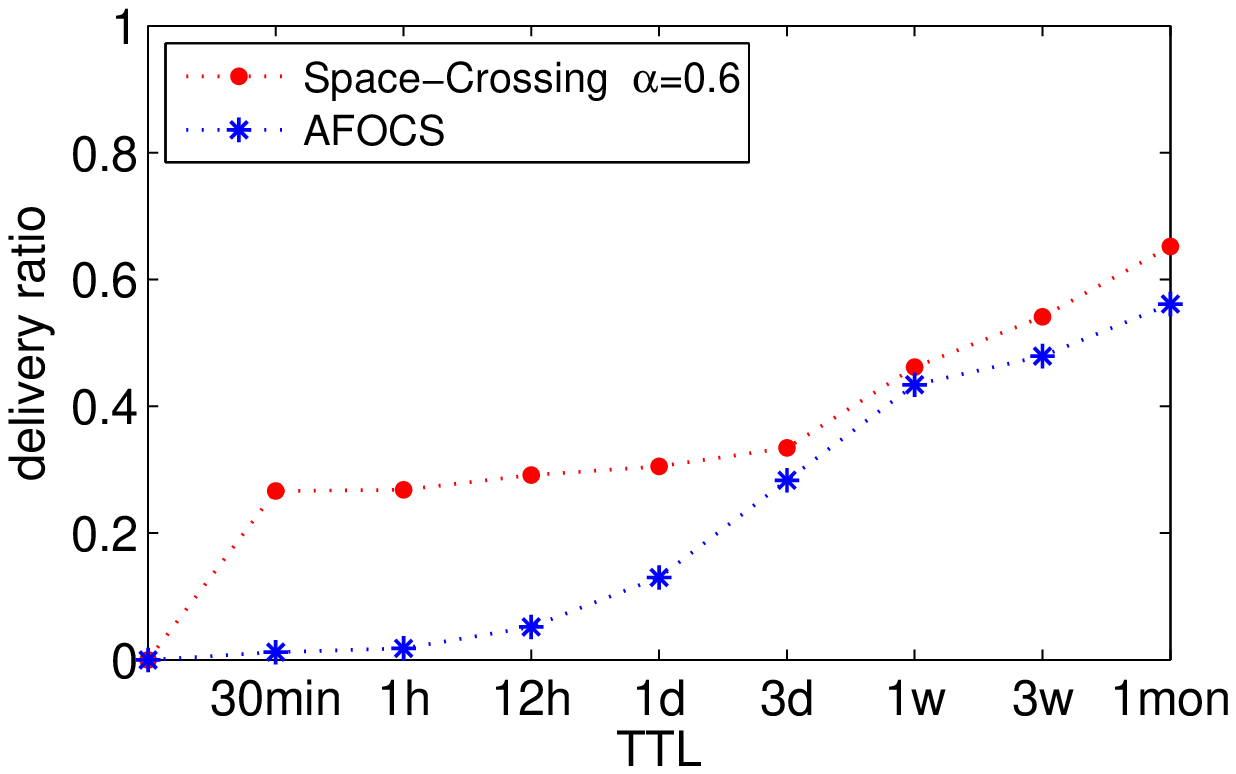}&
\hspace {-0.2in}
\includegraphics[width=0.55\columnwidth]{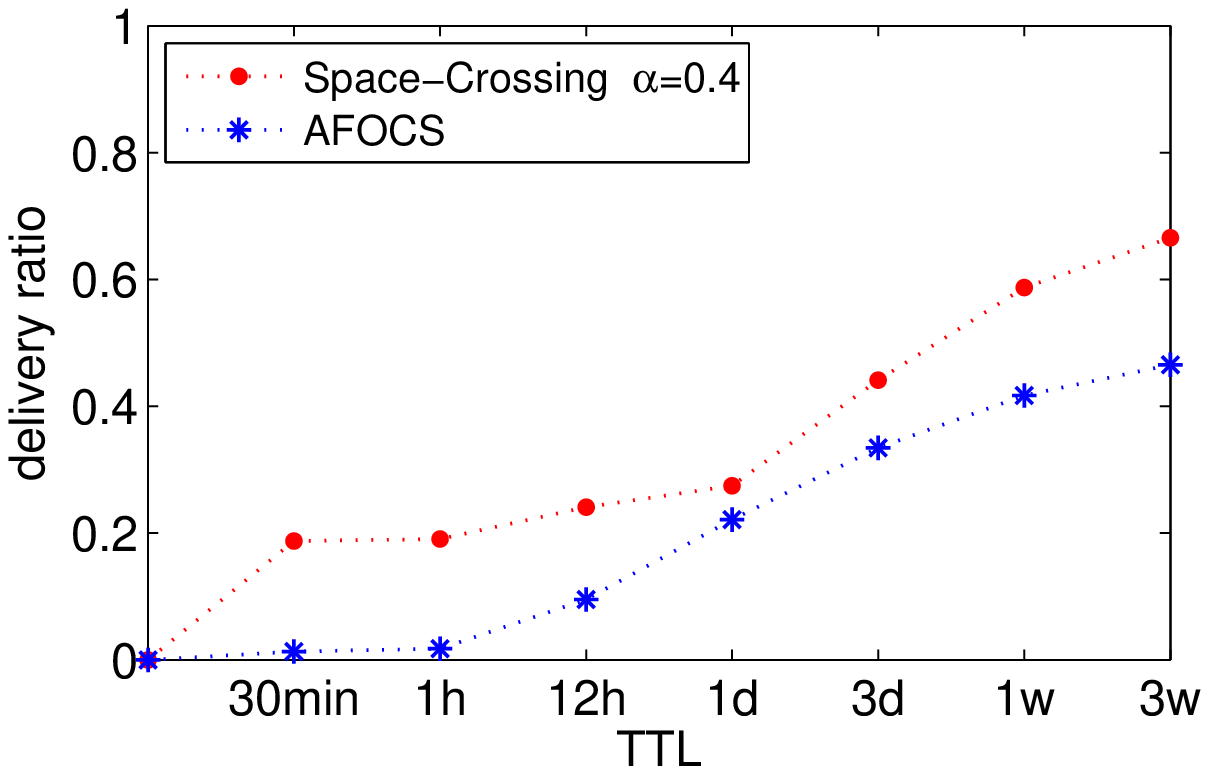}\\
(a) & (b)
\end{tabular}
\vspace {-0.1in}
\caption{The role of space-crossing community. Fig(a) is based on MIT Reality Mining dataset and Fig(b) is based on UIM dataset.} \label{role}
\end{center}
\vspace{-0.1in}
\end{figure}
 In SAAS algorithm, when calculating  \emph{pearson social similarity}, we require the community detection results of  the current social network. Here, especially, we use space-crossing community detection results and AFOCS \cite{overlap} detection results respectively to validate the role of space-crossing community.

From Fig.\ref{role} (a) and Fig.\ref{role} (b), we can see, in both MIT Reality Mining and UIM datasets, SAAS using the space-crossing community detection shows  better performance than the SAAS using  AFOCS  detection  in data forwarding. It  validates the important role of space-crossing community in data forwarding for MSNs.

From above results and analysis, SAAS has proved its competitive ability. In nature, it benefits from the strong communication community (SC Community) brought about by APs. According to SC Communities and node local activity, nodes can choose the appropriate relays to achieve a high efficient data forwarding.

\section{Discussion} \label{Discussion}
\subsection{The Detection Goodness of Space-Crossing Community}

In Section \ref{space-crossing}, we give a space-crossing community detection method. However, so far, there exist several benchmarks \cite{girvan2002community,lancichinetti2008benchmark,lancichinetti2009benchmarks} only for evaluating the goodness of the traditional community detection, not fit for our space-crossing community. From another perspective, our space-crossing community is a new community structure for fully utilizing APs to improve data forwarding  in MSNs. The evaluation of detection goodness is not the main topic of this paper.

\subsection{The Threshold Value for Criterion $\mathcal{S}^{a}$}

The threshold value $\alpha$ is related to the dataset. In above experiments, we can see that the dense networks prefer a relatively larger $\alpha$, while the sparse networks prefer a relatively smaller $\alpha$. However, there exist other factors which will influence the parameter $\alpha$. For example, in the dataset, if  the number of overlapped PP and AP Communities is small, the density sensitivity of $\alpha$ will decrease. If the degree of overlapped PP and AP Communities are fixed, the density sensitivity of $\alpha$ will disappear.
\section{Related Work} \label{Related Work}

In this paper, we propose a Space-Crossing Community Detection method for the hybrid AP underlying infrastructure and study its impact on data forwarding in Mobile Social Networks (MSNs).

For community detection, there have existed many classical centralized algorithms that are applied in the area of social networks, biological networks, commercial networks and so on. The recent reviews \cite{fortunato2010community} and \cite{lancichinetti2009community} may serve as introductory reading in this domain.
In the pioneering work, Newman and Girvan \cite{newman2004finding} constructed communities by removing links iteratively based on the betweenness value. The concept of MODULARITY $Q$ was given to estimate the goodness of a community partition. Then, Newman \cite{newman2004analysis} extended the work to weighted community detection and  Leicht \etal \cite{leicht2008community} transformed the undirected community detection problem to the directed one through importing the transposed matrix. Based on MODULARITY $Q$, many optimized algorithms were proposed \cite{clauset2004finding,guimera2005functional,blondel2008fast,rosvall2008maps}. Besides, as a milestone work, Palla \etal \cite{palla2005uncovering}  proposed a K-CLIQUE method to address the overlapped problem in community detection. As the development of mobile networks, the dynamic problem has to be settled. Some algorithms were proposed for this, such as Particle-And-Density \cite{kim2009particle} and QCA \cite{nguyen2011adaptive}. However,  above centralized community detection methods cost high in computation and difficult to implement in a
distributed ad hoc manner for MSNs. Thus,  Hui's  distributed community detection method  \cite{hui2007distributed} and  AFOCS \cite{overlap} method (a decentralized detection method using local information to tackle network changes) come into being. Nevertheless, so far, both the centralized and distributed methods aim at the non-organized distributed underlying infrastructure. They ignore the hybrid AP organization way in reality. From new perspective, we give the concept and the detection method of the space-crossing community brought about by this hybrid structure.


For data forwarding, some studies have shown that exploiting social relationships can achieve better data forwarding performances.
Daly and Haahr \cite{daly2007social} proposed SimBet  forwarding algorithm in Delay Tolerant Networks (DTNs). It used betweenness centrality and social similarity to increase the probability of a successful data forwarding.
Hui \etal \cite{hui2008bubble} proposed an algorithm called BUBBLE RAP in DTNs, with making use of node centrality and weighted k-clique community structure to enhance delivery performance.
Gao \etal \cite{gao2009multicasting} studied multicast in DTNs from the social network perspective.
Fan et al. \cite{fan2012geo} studied a geo-community-based broadcasting scheme for mobile social
networks by exploiting node geo-centrality and geo-community.
Nguyen \etal \cite{overlap} proposed a community-based data forwarding algorithm called Nguyen's Routing, by using the number of common interests as forwarding criterion.
Wu \etal \cite{wu2013homing} proposed a community home-based multi-copy routing scheme in MSNs.
However, none considers the positive role of space-crossing brought about by the hybrid underlying network with APs support in data forwarding.


To the best of our knowledge, this
is the first paper that studies the space-crossing community detection and its impact on data forwarding in MSNs by taking the hybrid underlying network with APs support into consideration.



\section{Conclusion} \label{Conclusion}

In this paper, we study a more realistic underlying infrastructure for MSNs, \ie, hybrid infrastructure with mobile users and APs. Due to the help of APs, this hybrid infrastructure brings about a new concept of community-space-crossing community. We give a space-crossing community detection method and propose a novel data forwarding  algorithm based on it called Similarity Attraction and AP Spreading (SAAS). SAAS utilizes
the detection results of space-crossing communities and node local activity to develop \emph{pearson social similarity} to
forward data quickly.
Simulation results show that space-crossing communities play an important
role in the data forwarding process.  SAAS achieves a better performance than existing social community-based algorithms. Our future work will
focus on  the relationship between the density of APs and  the performance of MSNs.



%
%
%

\bibliographystyle{IEEEtran}
\bibliography{MSN}

\begin{thebibliography}{10}
\providecommand{\url}[1]{#1}
\csname url@samestyle\endcsname
\providecommand{\newblock}{\relax}
\providecommand{\bibinfo}[2]{#2}
\providecommand{\BIBentrySTDinterwordspacing}{\spaceskip=0pt\relax}
\providecommand{\BIBentryALTinterwordstretchfactor}{4}
\providecommand{\BIBentryALTinterwordspacing}{\spaceskip=\fontdimen2\font plus
\BIBentryALTinterwordstretchfactor\fontdimen3\font minus
  \fontdimen4\font\relax}
\providecommand{\BIBforeignlanguage}[2]{{%
\expandafter\ifx\csname l@#1\endcsname\relax
\typeout{** WARNING: IEEEtran.bst: No hyphenation pattern has been}%
\typeout{** loaded for the language `#1'. Using the pattern for}%
\typeout{** the default language instead.}%
\else
\language=\csname l@#1\endcsname
\fi
#2}}
\providecommand{\BIBdecl}{\relax}
\BIBdecl

\bibitem{pietilainen2009mobiclique}
A.-K. Pietil{\"a}inen, E.~Oliver, J.~LeBrun, G.~Varghese, and C.~Diot,
  ``Mobiclique: middleware for mobile social networking,'' in \emph{Proc. ACM
  WOSN 2009}.

\bibitem{Foursquare}
``Foursquare.''\hskip 1em plus 0.5em minus 0.4em\relax http://foursquare.com.

\bibitem{yang2010smalltalker}
Z.~Yang, B.~Zhang, J.~Dai, A.~Champion, D.~Xuan, and D.~Li, ``E-smalltalker: A
  distributed mobile system for social networking in physical proximity,'' in
  \emph{Proc. IEEE ICDCS 2010}.

\bibitem{sony}
``Sony ps vita-near.''\hskip 1em plus 0.5em minus 0.4em\relax
  http://us.playstation.com/psvita.

\bibitem{hui2008bubble}
P.~Hui, J.~Crowcroft, and E.~Yoneki, ``Bubble rap: social-based forwarding in
  delay tolerant networks,'' in \emph{Proc. ACM MobiHoc 2008}.

\bibitem{gao2009multicasting}
W.~Gao, Q.~Li, B.~Zhao, and G.~Cao, ``Multicasting in delay tolerant networks:
  a social network perspective,'' in \emph{Proc. ACM MobiHoc 2009}.

\bibitem{fan2012geo}
J.~Fan, J.~Chen, Y.~Du, W.~Gao, J.~Wu, and Y.~Sun, ``Geo-community-based
  broadcasting for data dissemination in mobile social networks,'' \emph{to
  appear in£º IEEE Trans. on Parallel and Distributed Systems}, 2012.

\bibitem{overlap}
N.~P. Nguyen, T.~N. Dinh, S.~Tokala, and M.~T. Thai, ``Overlapping communities
  in dynamic networks: their detection and moibile applications,'' in
  \emph{Proc. ACM MobiCom 2011}.

\bibitem{girvan2002community}
M.~Girvan and M.~E. Newman, ``Community structure in social and biological
  networks,'' \emph{Proceedings of the National Academy of Sciences}, vol.~99,
  no.~12, pp. 7821--7826, 2002.

\bibitem{porter2009communities}
M.~Porter, J.~Onnela, and P.~Mucha, ``Communities in networks,'' \emph{Notices
  of the AMS}, vol.~56, no.~9, pp. 1082--1097, 2009.

\bibitem{fortunato2010community}
S.~Fortunato, ``Community detection in graphs,'' \emph{Physics Reports}, vol.
  486, no. 3-5, pp. 75--174, 2010.

\bibitem{uiuc-uim-2012-01-24}
K.~Nahrstedt and L.~Vu, ``{CRAWDAD} data set uiuc/uim (v. 2012-01-24),''
  Downloaded from http://crawdad.cs.dartmouth.edu/uiuc/uim, Jan. 2012.

\bibitem{eagle2009inferring}
N.~Eagle, A.~Pentland, and D.~Lazer, ``Inferring friendship network structure
  by using mobile phone data,'' \emph{Proceedings of the National Academy of
  Sciences}, vol. 106, no.~36, pp. 15\,274--15\,278, 2009.

\bibitem{hossmann2010know}
T.~Hossmann, T.~Spyropoulos, and F.~Legendre, ``Know thy neighbor: Towards
  optimal mapping of contacts to social graphs for dtn routing,'' in
  \emph{Proc. IEEE INFOCOM 2010}.

\bibitem{newman2004finding}
M.~Newman and M.~Girvan, ``Finding and evaluating community structure in
  networks,'' \emph{Physical Review E}, vol.~69, no.~2, p. 026113, 2004.

\bibitem{newman2004analysis}
M.~Newman, ``Analysis of weighted networks,'' \emph{Physical Review E},
  vol.~70, no.~5, p. 056131, 2004.

\bibitem{clauset2004finding}
A.~Clauset, M.~Newman, and C.~Moore, ``Finding community structure in very
  large networks,'' \emph{Physical Review E}, vol.~70, no.~6, p. 066111, 2004.

\bibitem{guimera2005functional}
R.~Guimera and L.~Amaral, ``Functional cartography of complex metabolic
  networks,'' \emph{Nature}, vol. 433, no. 7028, pp. 895--900, 2005.

\bibitem{blondel2008fast}
V.~Blondel, J.~Guillaume, R.~Lambiotte, and E.~Lefebvre, ``Fast unfolding of
  communities in large networks,'' \emph{Journal of Statistical Mechanics:
  Theory and Experiment}, vol. 2008, p. P10008, 2008.

\bibitem{rosvall2008maps}
M.~Rosvall and C.~Bergstrom, ``Maps of random walks on complex networks reveal
  community structure,'' \emph{Proceedings of the National Academy of
  Sciences}, vol. 105, no.~4, p. 1118, 2008.

\bibitem{leicht2008community}
E.~Leicht and M.~Newman, ``Community structure in directed networks,''
  \emph{Physical Review Letters}, vol. 100, no.~11, p. 118703, 2008.

\bibitem{palla2005uncovering}
G.~Palla, I.~Derenyi, I.~Farkas, and T.~Vicsek, ``Uncovering the overlapping
  community structure of complex networks in nature and society,''
  \emph{Nature}, vol. 435, no. 7043, pp. 814--818, 2005.

\bibitem{gregory2010finding}
S.~Gregory, ``Finding overlapping communities in networks by label
  propagation,'' \emph{New Journal of Physics}, vol.~12, no.~10, p. 103018,
  2010.

\bibitem{khadivi2011network}
A.~Khadivi, A.~Rad, and M.~Hasler, ``Network community-detection enhancement by
  proper weighting,'' \emph{Physical Review E}, vol.~83, no.~4, p. 046104,
  2011.

\bibitem{daly2007social}
E.~Daly and M.~Haahr, ``Social network analysis for routing in disconnected
  delay-tolerant manets,'' in \emph{Proc. ACM MobiHoc 2007}.

\bibitem{growingnetworks}
F.~Papadopoulos, M.~Kitsak, M.~{\'A}. Serrano, M.~Bogu{\~n}{\'a}, and
  D.~Krioukov, ``Popularity versus similarity in growing networks,''
  \emph{Nature}, vol. 489, pp. 537--540, 2012.

\bibitem{wu2012social}
J.~Wu and Y.~Wang, ``Social feature-based multi-path routing in delay tolerant
  networks,'' in \emph{Proc. IEEE INFOCOM 2012}.

\bibitem{keranen2009one}
A.~Ker{\"a}nen, J.~Ott, and T.~K{\"a}rkk{\"a}inen, ``The one simulator for
  {DTN} protocol evaluation,'' in \emph{Proc. ICST SIMUTools 2009}.

\bibitem{lancichinetti2008benchmark}
A.~Lancichinetti, S.~Fortunato, and F.~Radicchi, ``Benchmark graphs for testing
  community detection algorithms,'' \emph{Physical Review E}, vol.~78, no.~4,
  p. 046110, 2008.

\bibitem{lancichinetti2009benchmarks}
A.~Lancichinetti and S.~Fortunato, ``Benchmarks for testing community detection
  algorithms on directed and weighted graphs with overlapping communities,''
  \emph{Physical Review E}, vol.~80, no.~1, p. 016118, 2009.

\bibitem{lancichinetti2009community}
------, ``Community detection algorithms: A comparative analysis,''
  \emph{Physical Review E}, vol.~80, no.~5, p. 056117, 2009.

\bibitem{kim2009particle}
M.-S. Kim and J.~Han, ``A particle-and-density based evolutionary clustering
  method for dynamic networks,'' \emph{Proceedings of the VLDB Endowment},
  vol.~2, no.~1, pp. 622--633, 2009.

\bibitem{nguyen2011adaptive}
N.~P. Nguyen, T.~N. Dinh, Y.~Xuan, and M.~T. Thai, ``Adaptive algorithms for
  detecting community structure in dynamic social networks,'' in \emph{Proc.
  IEEE INFOCOM 2011}.

\bibitem{hui2007distributed}
P.~Hui, E.~Yoneki, S.~Y. Chan, and J.~Crowcroft, ``Distributed community
  detection in delay tolerant networks,'' in \emph{Proc. ACM MobiArch 2007}.

\bibitem{wu2013homing}
J.~Wu, M.~Xiao, and L.~Huang, ``Homing spread: Community home-based multi-copy
  routing in mobile social networks,'' in \emph{Proc. IEEE INFOCOM 2013}.

\end{thebibliography}

\end{document}